Genome-Guided Interpretable Screening of Phase-Stable, Lead-Free Double Perovskite Absorbers for All-Inorganic Semiconductors, Sensors, and Photovoltaics with DFT-Validated Design Rules

Nafis Ahtasum [1,2,*], Sohanur Rahman Sohan[1,3], Md. Mostaq Ahmed Himel[1,2], Md. Zahid Hassan[3], Muhammad Harussani Moklis[1,4,5], Masud Rana Rashel[6,7], Hasan Jamil[8], AKM Kamrul Islam[9], Mouhaydine Tlemcani[6]

[1]Center for Material, Climate and Energy, Research and Analysis Institute, RAI Initiative Ltd, Dhaka, Bangladesh

[2]Department of Apparel Engineering, Bangladesh University of Textiles, Dhaka-1208, Bangladesh

[3]Department of Textile Engineering Management, Bangladesh University of Textiles, Dhaka-1208, Bangladesh

[4]Department of Chemical and Environmental Engineering, Faculty of Engineering, Universiti Putra Malaysia, Serdang 43400, Selangor, Malaysia

[5]Energy Science and Engineering, Department of Transdisciplinary Science and Engineering, Institute of Science Tokyo, 2-12-1, Ookayama, Meguro-ku, Tokyo 152-8550, Japan

[6]Instrumentation and Control Laboratory, Center for sci-tech Research in Earth System and Energy, University of Évora, Portugal

[7]Department of Computer Science and Engineering, Daffodil International University, Bangladesh

[8]Department of Computer Science, University of Idaho, 875 Perimeter Drive, Moscow, ID 83844, USA

[9]Computational Science and Engineering, North Carolina A&T State University, Greensboro, North Carolina, USA

*=corresponding author (email= nafisahtasum666@gmail.com)

Abstract

The discovery of stable, lead-free halide perovskites for optoelectronic applications remains constrained by the vast compositional space and the limited physical interpretability of conventional data-driven screening approaches. Here, we present a genome-guided, physics-informed screening framework that decodes thermodynamic stability and optoelectronic behavior through four physically interpretable descriptor families–packing, bonding, polarization, and electronic identity. Trained on 1,221 DFT-calculated $A_2BB'X_6$ compounds, machine-learning surrogates achieve robust predictive performance, with a recall-optimized stability classifier (ROC–AUC = 0.92) and an XGBoost regressor for band-gap prediction ($R^2$ = 0.93 on held-out test data). Application of a staged inverse-design constraint stack to 13,088 charge-balanced, lead-free compositions reduces the search space to five DFT-validated, phase-stable semiconductors: $Rb_2SnMnBr_6$, $Cs_2CdSnBr_6$, $Cs_2CdSnI_6$, $Cs_2KGaI_6$, and $Cs_2AgAlBr_6$. These candidates lie on the convex hull ($E_hull \leq 0$ meV atom$^{-1}$), preserve ordered double-perovskite structures, and exhibit strong optical absorption ($\alpha_peak \approx 10^5$ cm$^{-1}$). Genotype–phenotype coupling analysis reveals a hierarchical control mechanism in which packing genes define the structural formability manifold, bonding genes govern near-edge optical transitions and conductivity, and Opto-electronic response genes regulate dielectric response and exciton screening ($\varepsilon_0$ = 4.6–8.2). Beyond candidate identification, this work establishes a generalizable paradigm for interpretable inverse design, where descriptor-level genomics is directly linked to experimentally relevant optoelectronic phenotypes. The resulting design rules provide a mechanistically grounded pathway for discovering stable, lead-free double perovskites for photovoltaics, sensing, and transparent electronic applications.

## 1. Introduction

All-inorganic halide perovskites are very popular for solar devices, photodetectors and other opto-electronic applications. The materials are interesting in that they have strong light absorption, tunable band gaps and have good charge transport properties. Most of the conventional halide perovskites, however, are composed of lead (Pb), which poses issues of toxicity and stability. For this reason, research is ongoing to find alternatives to lead. One of the most promising is the class of so-called “halide double perovskites” $A_2BB'X_6$, where A and B are monatomic elements and B′ is a bivalent element such as lead. There are several different forms of cation substitution that can provide charge balance and also generate a wide variety of material combinations. Such materials have been recently reported to possess good structural diversity, stability and tunable optoelectronic properties [1,2].

Nevertheless, it is not immediately obvious how to find suitable double perovskites where the lead is replaced. There are a wide number of possible compositions of type $A_2BB'X_6$ but only a few compounds have been prepared or investigated in detail. Another difficulty is that the properties that are required are very near to one another. The material could also have a good band gap but experience poor stability, weak optical absorption and weak charge transport. It was noted recently that simultaneously improving the properties of structural stability, thermodynamic stability, band-gap position, and optoelectronic properties is challenging in the case of lead-free perovskite materials [1–3].

To accelerate this search, high-throughput density functional theory (DFT) and machine learning (ML) have become increasingly important in recent years [4]. Earlier studies showed that data-driven models can predict important quantities such as thermodynamic stability, formation energy, and electronic band gap across broad perovskite spaces with useful accuracy. For example, Landini [5]trained a neural-network model for band-gap prediction and screened 7,056 lead-free halide double perovskites, followed by hybrid-DFT evaluation of a reduced subset. Chen [6] developed ML models using a dataset containing 3,720 $ABX_3$ perovskites and 2,660 double perovskites, targeting both band gap and formation energy while also emphasizing interpretability through feature-attribution analysis. In parallel, the high computational cost of first-principles methods has often limited purely DFT-based exploration to smaller candidate sets; for example, Gao [7] performed high-throughput DFT screening of 760 $Cs_2B^{2+}B'^{2+}X_6$ compositions for lead-free photovoltaic absorbers. These studies clearly show that both ML-based and DFT-based screening have greatly accelerated lead-free perovskite discovery.

However, a key limitation of many existing data-driven workflows is that they are optimized mainly for predictive efficiency rather than physical interpretability. In practice, such models often provide ranked candidate lists without clearly showing which chemical variables should be tuned to improve structural formability, framework stability, dielectric constant, or optical transition strength. This issue is now widely recognized in the broader materials-ML literature. Recent reviews on knowledge-driven, physics-informed, and interpretable machine learning have called for the next generation of computational materials discovery to go beyond predictive accuracy to provide transferable design principles, insight into mechanism, and chemically meaningful control variables [1,8,9]. This is particularly crucial in the context of lead-free double perovskites, where a slight variation in the size, electronegativity contrast, polarizability and the character of the band-edge orbitals can significantly impact the stability and functionality of the material functionality [8,10–12].

For this class of materials, a useful discovery framework should therefore do more than screen quickly. It should distinguish structural feasibility from downstream property optimization, retain descriptors with clear chemical meaning, and reveal how composition-level variables connect to observable optoelectronic behavior. In this sense, descriptor engineering should not be treated only as a numerical preprocessing step, but as a physically motivated representation of the chemistry governing perovskite formability, energetic accessibility, and functional response.

Here, we present a genome-guided, interpretable screening framework for lead-free halide double perovskites. In this approach, each $A_2BB'X_6$ composition is encoded using a compact set of physically motivated descriptors, referred to here as a stability genome, that captures four broad aspects of materials chemistry: geometric packing, bonding environment, polarization response, and electronic identity. These descriptors are used to train ML surrogates for thermodynamic stability classification and band-gap prediction, and the resulting models are applied to a chemically constrained library of charge-balanced lead-free compositions. A staged screening strategy combining ML inference, geometry-based filtering, band-gap targeting, and deployability constraints is then used to reduce the broad chemical space to a small set of DFT-actionable candidates.

The significance of the present work lies not only in candidate identification, but also in the extraction of validated design rules from an interpretable descriptor framework. By combining machine-learning screening with targeted first-principles validation, we connect descriptor modules to structural formability, thermodynamic stability, dielectric constant, and optical response. In this way, the present study moves beyond black-box ranking and establishes a genome-guided, interpretable screening strategy for navigating the chemistry of lead-free double perovskites. More broadly, it shows how physically organized descriptor spaces can transform high-throughput screening from a purely predictive exercise into a design-oriented framework for discovering and understanding next-generation lead-free perovskite materials.

## 2. Methods: Genome-guided interpretable screening framework

### 2.1 Workflow overview

We developed a genome-guided inverse-design workflow to identify phase-stable, lead-free double perovskite absorbers with targeted optoelectronic properties[13]. Rather than starting from known materials and screening them forward, the workflow begins from desired performance windows and maps them back to composition through physically interpretable descriptors.

The workflow consists of five connected steps: (i) enumeration of a large charge-balanced $A_2BB'X_6$ chemical space, (ii) encoding of each composition into descriptor-based genome features, (iii) training of machine-learning surrogate models for stability and band-gap prediction, (iv) application of a staged inverse-design screening strategy based on physical and practical constraints, and (v) targeted DFT validation of the shortlisted candidates[14,15]. The pipeline, illustrated in Figure 1, combines chemical-space enumeration, descriptor-based genome encoding, evolutionary-optimized machine-learning surrogate models, a staged inverse-design constraint stack, and DFT phenotype closure for structural, electronic, dielectric, optical, and transport-related validation.

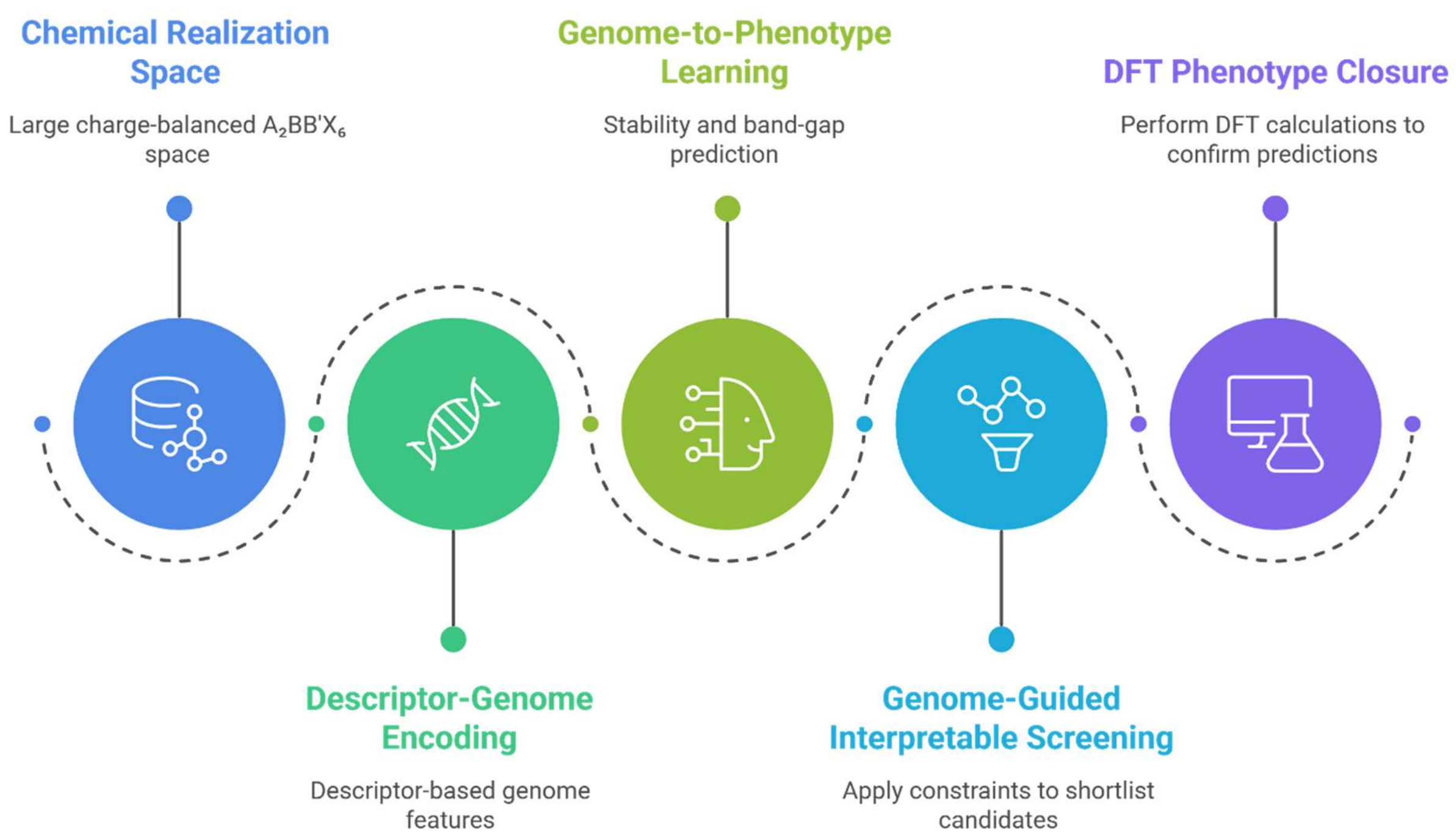

*Figure 1. Genome-guided interpretable screening workflow used in this study.*

## 2.2 Chemical-space construction and structural filtering

A lead-free, all-inorganic $A_2BB'X_6$ double-perovskite chemical space was constructed by combining explicit charge neutrality with structural rules consistent with corner-sharing $BX_6$ octahedra and rock-salt ordering of the B/B′ sites. Two valence families were considered: heterovalent $A_2B^{+}B'^{3+}X_6$ and homovalent $A_2B^{2+}B'^{2+}X_6$. The element pools were defined as A = Li, Na, K, Rb, Cs, Cu, Ag, and Au; B/B′ = alkaline earths (Be–Ba), selected transition metals (Mn–Zn, Pd, Ir, Pt), and post-transition or metalloid elements (Al, Ga, In, Sn, Ge, Sb, Bi, Te); and X = F, Cl, Br, and I. Pb-containing compositions were excluded by design.

All possible formulas were generated by exhaustive combinatorial enumeration using a Python workflow based on *pymatgen*. A composition was retained only if it satisfied strict charge neutrality:

$$2q_A + q_B + q_{B'} + 6q_X = 0$$

where $q_A$, $q_B$, $q_{B'}$, and $q_X$ are the nominal oxidation states of the corresponding elements. For the homovalent family, $B/B'$ combinations were canonicalized to avoid duplicated formulas caused by permutation. In total, this procedure produced 13,088 unique, charge-balanced compositions[16].

2.3 DFT-labeled dataset and target definitions

To train the machine-learning models, we assembled a DFT-labeled dataset of 1,221 unique $A_2BB'X_6$ halide double perovskites using a reproducible Python-based workflow built on Materials Project data (Table 1). Thermodynamic stability was quantified using the energy above the convex hull ($E_{hull}$), which is widely used in high-throughput screening as a descriptor of 0 K phase stability. Compounds with $E_{hull} \leq 25$ meV atom$^{-1}$ were labeled as stable, whereas those with $E_{hull} > 25$ meV atom$^{-1}$ were labeled as unstable[17]. This binary target was used for stability classification. The dataset contains 401 homovalent compounds with $B^{2+}/B'^{2+}$ cation pairs and 820 heterovalent compounds with $B^{+}/B'^{3+}$ cation pairs, allowing the model to learn across both oxidation-state families. Table 1 summarizes the DFT dataset, target definitions, and data split used for ML model development.

*Table 1. DFT-labeled dataset summary and target definitions*

| Property | Description |
|---|---|
| System | $A_2BB'X_6$ double perovskites[18] |
| Dataset size | 1,221 |
| Source | Materials Project |
| Composition types | 401 homovalent; 820 heterovalent |
| Stability metric | E_hull (meV atom$^{-1}$) |
| Stability criterion | ≤ 25 (stable); > 25 (unstable) |
| Classification target | Stability label |
| Regression target | PBE band gap (E_g) |
| Data split | 80:20 (train:test), stratified[19] |
| Data content | E_hull, E_g, relaxed structures |

In addition to stability, the scalar-relativistic PBE electronic band gap (E_g) was used as a continuous regression target. For model development, the dataset was divided into training (80%) and testing (20%) subsets using stratified sampling based on the stability label. We note that $E_{hull}$-based labeling reflects thermodynamic stability at 0 K and does not explicitly account for finite-temperature effects, anharmonic lattice dynamics, moisture or oxygen sensitivity, or other degradation pathways that are particularly relevant to halide perovskites. Therefore, in this work, the term "stable" should be interpreted as DFT-predicted thermodynamic stability rather than guaranteed experimental or ambient-condition stability.

### 2.4 Stability-genome descriptor design

Each $A_2BB'X_6$ composition was represented using a set of physically interpretable descriptors, referred to here as the stability genome. These descriptors were designed to capture the dominant chemical and structural factors governing thermodynamic stability, band-gap behavior, dielectric constant, and optical response.

i. Packing genes comprise ionic radii and tolerance factors (t, μ, τ), which describe geometric compatibility, octahedral fit, and structural formability. Bonding genes include bond-strength proxies and electronegativity differences, which reflect framework cohesion, bond polarity, and metal–halide hybridization. The properties of atomic polarizability, ionization energy and electron affinity are examples of Opto-electronic response genes that have an influence on the dielectric response, charge transfer and screening behavior.

ii. Periodic-trend and band-edge chemistry effects were also preserved using electronic-identity descriptors like the valence electron count and atomic number.

iii. Electronic-identity descriptors, such as valence electron count and atomic number, were also retained to capture band-edge chemistry and periodic-trend effects. Starting from a larger pool of candidate variables, redundant and weakly informative features were removed using correlation analysis and feature-relevance criteria[20–26].

iv. The final descriptor set contains 13 low-collinearity and physically meaningful features. All descriptors were standardized to zero mean and unit variance prior to model training.

### 2.5 Machine-learning surrogates and evolutionary hyperparameter optimization

To enable efficient screening of the large compositional space, we developed machine-learning surrogate models for two supervised learning tasks: (i) classification of thermodynamic stability and (ii) prediction of the scalar-relativistic PBE band gap ($E_g$). The classification models were trained to predict the DFT-derived stability label defined by the $E_{\mathrm{hull}}$-based criterion described in Section 2.3, while the regression models were trained to predict the corresponding PBE band-gap values.

For thermodynamic stability classification, three algorithms were examined, namely Decision Tree (DT), Random Forest (RF), and extreme gradient boosting (XGBoost). For band-gap prediction, Random Forest regression, support vector regression (SVR) with a radial basis function kernel, and XGBoost regression were evaluated[27–29]. All models were trained using the DFT-labeled dataset containing 1,221 $A_2BB'X_6$ compositions and were then used to screen the full enumerated library of 13,088 compositions.

The DFT-labeled dataset was assembled independently from the enumerated compositional library using a separate Python-based workflow built on Materials Project data. Thus, the model training set and the screened candidate space were generated independently, although both belong to the same $A_2BB'X_6$ halide double-perovskite family. The trained models were therefore used to identify and prioritize promising compositions within the enumerated design space based on patterns learned from the independently collected DFT dataset.

To improve model performance and avoid empirical manual tuning, hyperparameter optimization was carried out using a genetic-algorithm-based evolutionary strategy. In this approach, an initial population of candidate hyperparameter combinations was generated randomly and then evolved through selection, crossover, and mutation. Optimization was done for 50 generations with a population size of 50 and model fitness was tested on the validation subset of the training data. The optimization objective for the stability-classification task was set to maximize the recall of the stable class, to minimize false-negative predictions when screening. In case of the band-gap regression task, the goal was to reduce the mean squared error (MSE).

Final optimized hyperparameters of the thermodynamic stability model is Decision Tree (max_depth = 6, min_samples_split = 5, min_samples_leaf = 9, criterion = gini, splitter = best, and max_features = 0.5699). The final model used for the prediction of the band gap was a regressor named XGBoost with the following hyperparameters: learning_rate = 0.0801, max_depth = 8, n_estimators = 380, subsample = 0.9114, reg_lambda = 0, reg_alpha = 0.3901, gamma = 0.3351, colsample_bylevel = 0.9534, and colsample_bytree = 0.9277. The optimized models were then retrained using the entire training set, and applied to the entire enumerated compositional library for high-throughput screening (HTS) (see Table S2).

2.6 Inverse-design screening strategy

A staged inverse-design constraint stack was implemented to translate target property requirements into candidate compositions. This hierarchical strategy enables systematic reduction of the large chemical space while preserving physically meaningful and application-relevant materials.

In Stage I, the ML stability classifier was used to identify compositions likely to be thermodynamically stable. In Stage II, geometry-based filtering using t, μ, and τ was applied to enforce structural compatibility with the double-perovskite framework. In Stage III, the band-gap regressor was used to select candidates within application-relevant target windows. In Stage IV, deployability restrictions have been strengthened and included chemistry-aware exclusion restrictions, as needed, in addition to structural admissibility restrictions. The final shortlisted compounds were tested using first principle DFT in Stage V[30]. The designed screening pipeline integrates three components: data-driven prediction, physically informed constraints, and allows the efficient navigation of the $A_2BB'X_6$ design space without sacrificing chemical plausibility and device relevance.

## 2.7 DFT framework, magnetism, and phenotype closure

First-principles calculations were performed in Materials Studio using $DMol^3$ within the spin-polarized GGA-PBE framework. Structural optimization was carried out using a DNP (3.5) basis set together with scalar-relativistic DSPP pseudopotentials for heavy elements. The geometry optimization was continued until the total-energy change became smaller than $1 \times 10^{-6}$ Ha and the residual forces were below 0.002 Ha $Å^{-1}$. Monkhorst-Pack k-point meshes were chosen based on convergence testing, so that the total energy was converged within a few meV $atom^{-1}$ and the band gap within ≤ 0.05 eV[31,32]. The final k-point grids used for the shortlisted compounds are provided in Section 2.8 and Table S6.

Optical properties, including the dielectric function, absorption coefficient, refractive index, reflectivity, loss function, and optical conductivity, were calculated using CASTEP with GGA-PBE and norm-conserving pseudopotentials under the same convergence conditions. Simulated powder XRD patterns were also generated to confirm the structural symmetry and phase purity of the selected compounds[31–33].

It should be mentioned that $DMol^3$ uses a numerical atomic orbital basis set instead of a plane-wave basis. In this work, $DMol^3$ was used to keep the computational workflow consistent for all candidate compounds within the same Materials Studio environment. Therefore, the calculated electronic properties were mainly used for comparative analysis and ranking within the same computational framework, rather than for direct comparison with results from other DFT codes.

Spin polarization was included for all systems. For compounds containing $Mn^{2+}$ and $Cu^{2+}$, the calculations were initialized in high-spin ferromagnetic states.

## 2.8 Convergence tests

Representative chlorides and bromides, with and without open-shell cations, were used to benchmark k-point meshes and basis-set quality. Parameters were increased until changes in total energies, lattice constants, and band gaps fell below a few meV $atom^{-1}$, ~0.01 Å, and ~0.05 eV, respectively. The same settings were adopted for optical-property calculations. Final convergence data and parameter sets are summarized in Table S4.

## 2.9 Effective mass and deformation-potential mobility

For the four DFT-validated double perovskites, carrier transport was assessed via effective masses and phonon-limited mobilities using the deformation-potential (DP) approach. Band structures were computed along Γ–X and Γ–L, and electron/hole effective masses were obtained by parabolic fits near the CBM and VBM. Directional values were averaged to yield scalar $m_e^*$ and $m_h^*$.

Mobilities were estimated using the 3D Bardeen–Shockley DP model [34],

$$\mu = \frac{2e\hbar^4 C}{3k_B T (m^*)^2 E_1^2},$$

with C(elastic constant) and $E_1$(deformation potential) extracted from total-energy and band-edge shifts under small uniaxial strains. Direction-resolved C, $E_1$, and $m^*$were combined to obtain $\mu_e$and $\mu_h$; direction-averaged values are used for discussion, with full anisotropic data relegated to the Supporting Information[35–38]. These transport metrics support the comparative assessment of the four compounds as candidate absorbers or transport layers.

## 3. Results and Discussion

### 3.1 Chemical space anatomy and formability manifold:

The genome-guided screening starts from a chemically broad $A_2BB'X_6$ compositional space. But it is physically constrained. By combining oxidation-state rules for heterovalent $A_2B^{+}B'^{3+}X_6$ and homovalent $A_2B^{2+}B'^{2+}X_6$ families, and removing duplicate formulas in the homovalent set, we obtained 13,088 unique, charge-balanced lead-free compositions (Figure 2). This construction was designed to keep large chemical diversity. It also maintained the realistic double-perovskite stoichiometry.

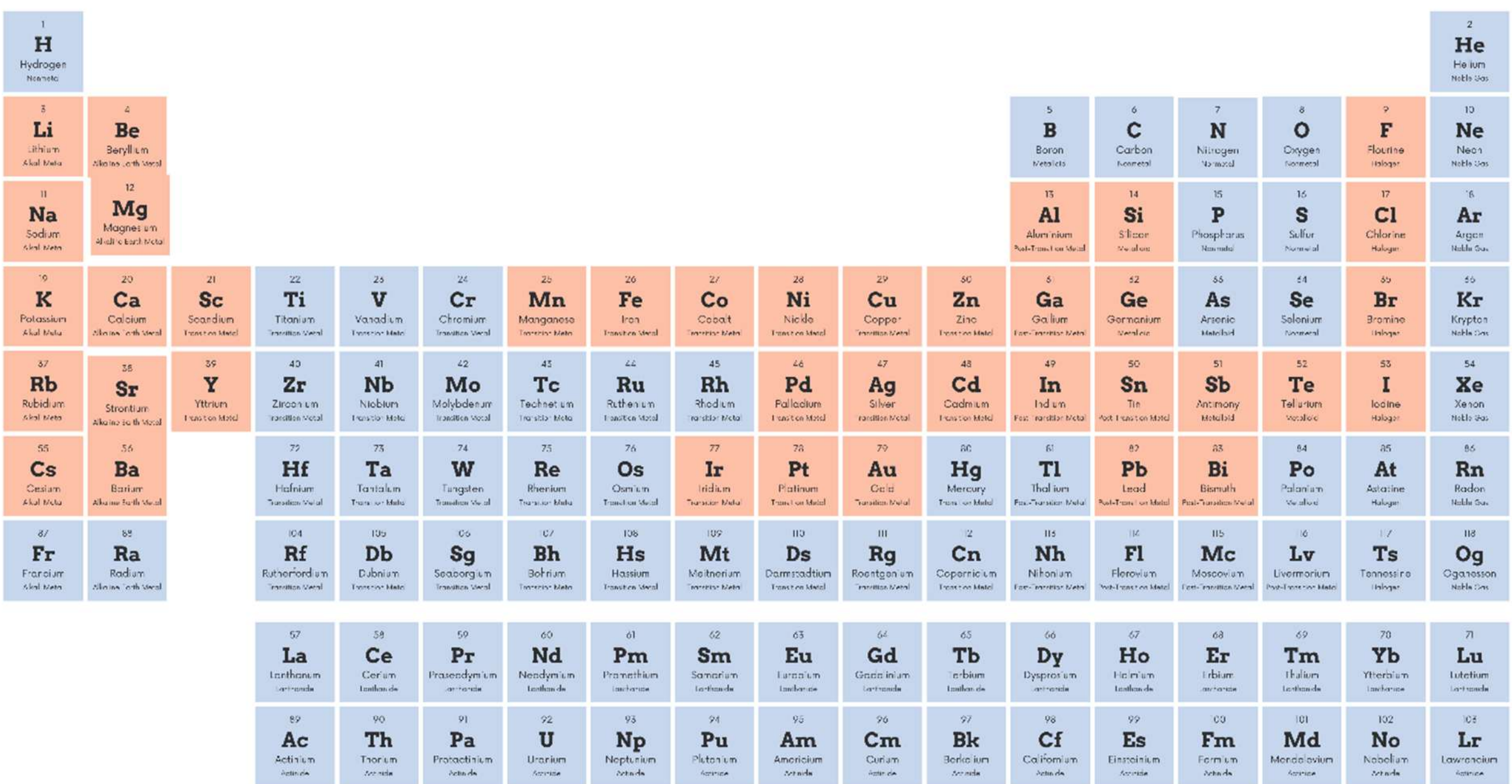

*Figure 2. Chemical-space anatomy. Element pools and oxidation-state families used to enumerate the $A_2BB'X_6$ library.*

Summary of the oxidation-state families, element pools, duplicate-handling rules, and geometric cutoffs used to construct the lead-free $A_2BB'X_6$ chemical library, as according to Table 2.

*Table 2. Chemical-space construction and formability criteria prior to ML screening.*

| Family | Oxidation-state rule | Element pools | Duplicate handling | Geometry cutoffs | Retained compositions |
|---|---|---|---|---|---|

| Heterovalent | $A_2B^{+}B'^{3+}X_6$ | A = Li, Na, K, Rb, Cs, Cu, Ag, Au; B/B′ = Be-Ba, Mn-Zn, Pd, Ir, Pt, Al, Ga, In, Sn, Ge, Sb, Bi, Te; X = F, Cl, Br, I; | None stated | 0.80 ≤ t ≤ 1.10; 0.414 ≤ μ ≤ 0.732; τ < 4.18 | Included in total library |
|---|---|---|---|---|---|
| Homovalent | $A_2B^{2+}B'^{2+}X_6$ | A = Li, Na, K, Rb, Cs, Cu, Ag, Au; B/B′ = Be-Ba, Mn-Zn, Pd, Ir, Pt, Al, Ga, In, Sn, Ge, Sb, Bi, Te; X = F, Cl, Br, I | Canonicalized to remove duplicate B/B′ formulas | 0.80 ≤ t ≤ 1.10; 0.414 ≤ μ ≤ 0.732; τ < 4.18 | Included in total library |
| Combined space | Charge-balanced $A_2BB'X_6$ | 8 A-site cations, 24 B/B′-site cations, 4 halides | Duplicate removal for homovalent set | 0.80 ≤ t ≤ 1.10; 0.414 ≤ μ ≤ 0.732; τ < 4.18 | 13,088 unique compositions |

Using Shannon ionic radii, we calculated the Goldschmidt tolerance factor ($t$), octahedral factor ($\mu$), and new tolerance factor ($\tau$) for each composition. The acceptance windows used in this work were $0.80 \leq t \leq 1.10$, $0.414 \leq \mu \leq 0.732$, and $\tau < 4.18$ (Table 2). These packing descriptors form the physical screening gate because they directly indicate steric compatibility and octahedral fit within the double-perovskite framework.

The chemical-space statistics show that the enumerated library is not uniformly distributed. The B–B′ combinatorial frequency map indicates higher density for several cation pairs involving $Bi^{3+}$, $Sb^{3+}$, $In^{3+}$, and $Ga^{3+}$ on the trivalent site, suggesting that oxidation-state constraints and ionic-size matching already shape the search space (Figure 3(a)). The geometric formability space further shows that only a limited region of the $\mu$-$t$ landscape falls inside the feasible manifold, confirming that many charge-balanced compositions remain structurally implausible (Figure 3(b)).

A clear halide trend is also observed in Figure 3(c). Chlorides (48.2%) and bromides (45.8%) show the highest fraction of compositions passing the packing criteria. But fluorides (12.5%) and iodides (15.4%) are remained much less.

This behavior reflects the effect of halide size on structural fit: $F^-$ tends to overconstrain cation packing, while $I^-$ often pushes the structure outside the optimum steric window. $Cl^-$ and $Br^-$, on the other hand, offer a more balanced geometric regime to stabilize the double-perovskite framework over a wider range of cations.

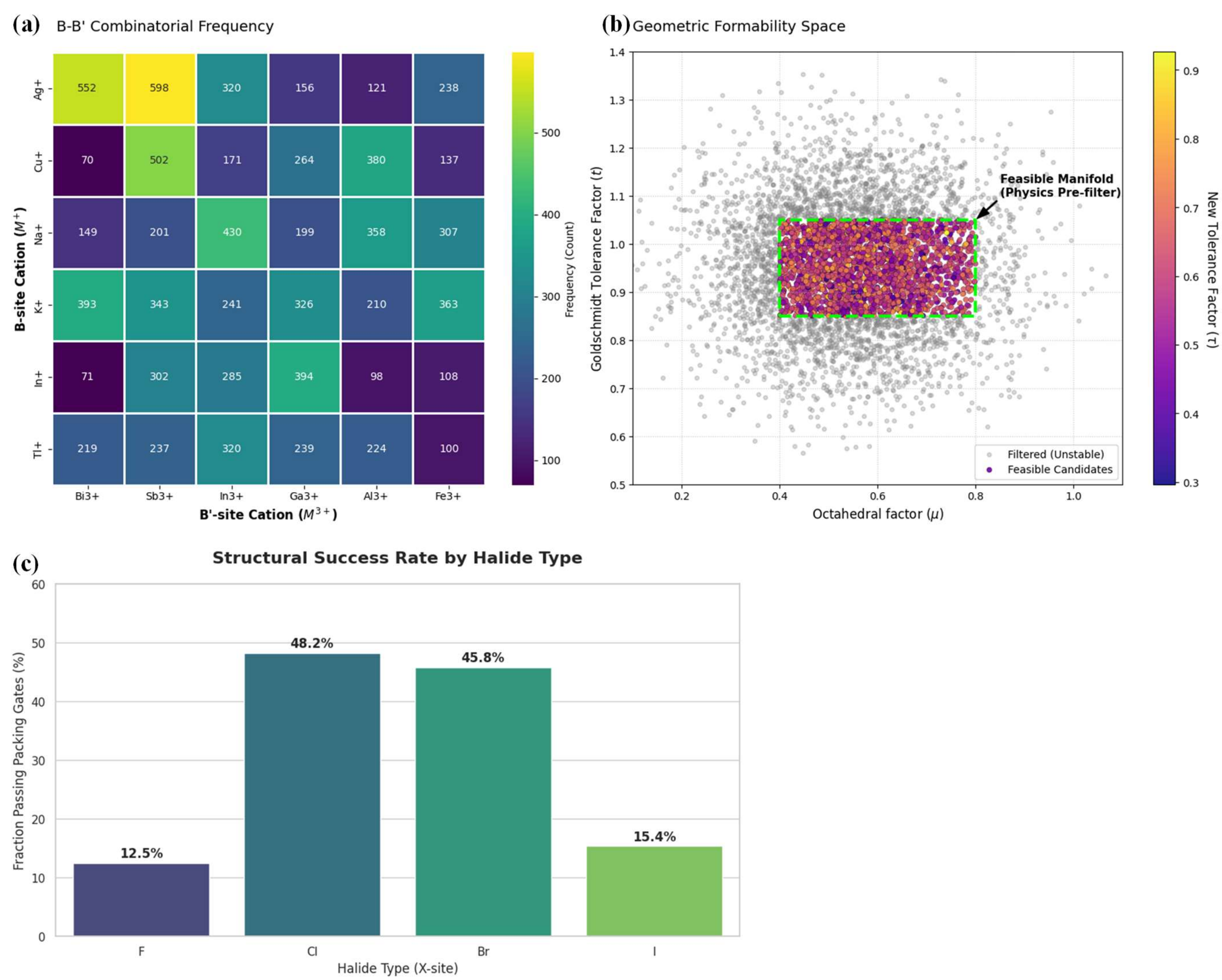

*Figure 3. Chemical-space formability statistics. (a) B–B' combinatorial density map aggregated overall A and X choices. (b) Distributions of $t$, $\mu$, and $\tau$ across the enumerated space, showing the formability manifold used as a physical screening gate. (c) Fraction of compositions retained after the geometry-based filters.*

Overall, these results define the chemical-space anatomy of the present discovery problem. Although the initial library is broad in composition, its physically meaningful region is much smaller and strongly shaped by halide chemistry and cation-size compatibility. The formability manifold therefore acts not only as a pre-filter, but also as the first interpretable layer of the genome-guided screening strategy.

## 3.2 Decoding the genome of stability: compact descriptor genes with physically interpretable roles

To improve the interpretability of the screening framework, we organized the descriptor space into a compact set of physically meaningful groups, which we refer to as the stability genome. We did not consider all descriptors as a general feature set in this work.  We designed this clusters to reflect the main chemical and structural factors. These factors are expected to influence double-perovskite formability, thermodynamic stability, and optoelectronic behavior, as represented in Figure 4.

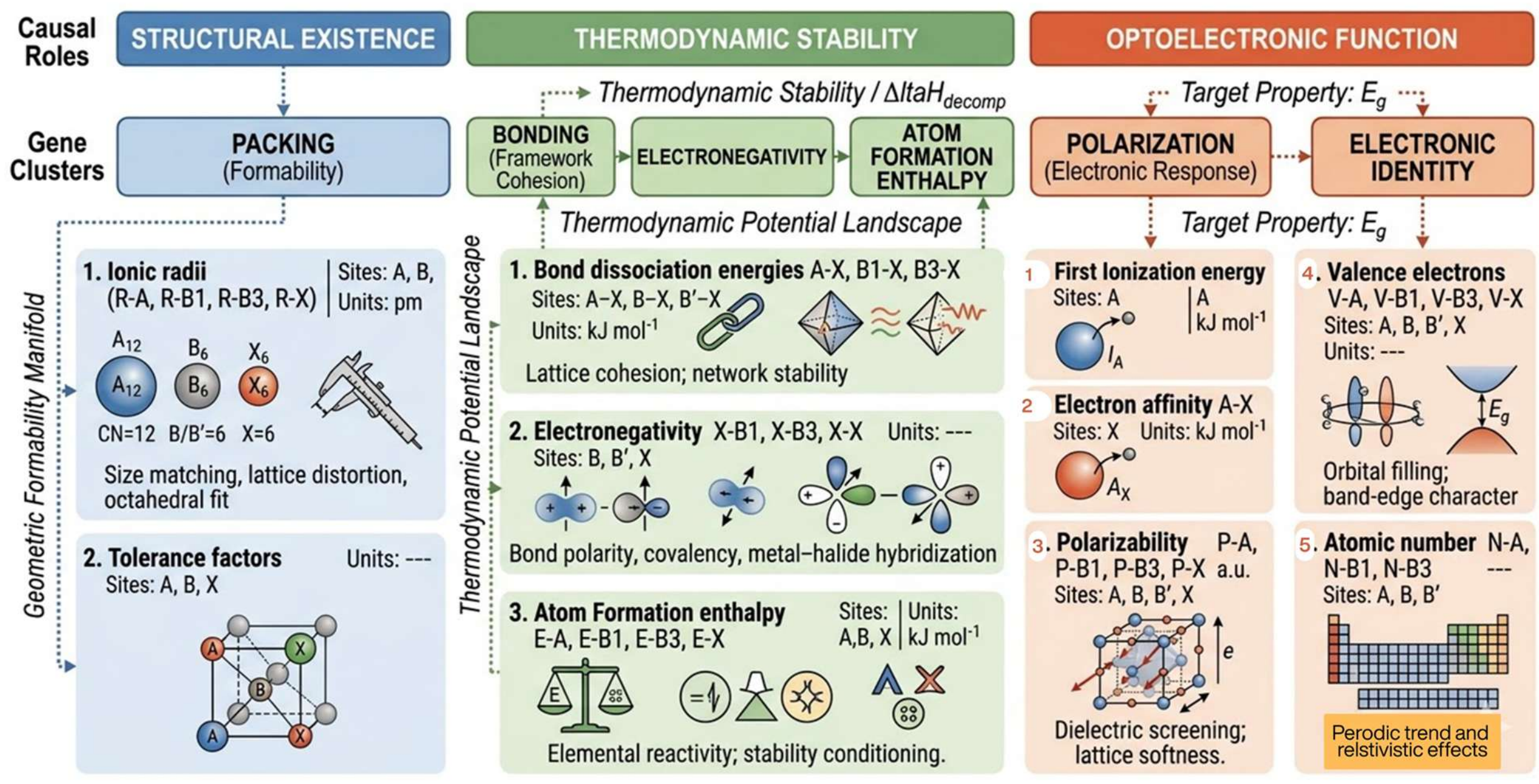

*Figure 4. Schematic illustration of the descriptor space clustered into packing, bonding, and optoelectronic-response categories.*

The first group, termed packing descriptors, describes structural existence through ionic radii and the formability metrics $t$, $\mu$, and $\tau$. These variables indicate directly, whether a given composition is geometrically compatible or not. The second group, bonding genes (descriptors), is connected with thermodynamic stability. It includes bond dissociation energies, electronegativity differences, and atomic formation enthalpies.

These descriptors were selected to represent framework cohesion, bond polarity, and the strength of metal-halide interactions. The third group, denoted as optoelectronic-response descriptors, includes ionization energy, electron affinity, polarizability, valence electron count, and atomic number. These variables retain information related to dielectric response, charge redistribution, periodic trends, and band-edge character, and are therefore relevant for optical and electronic screening.

Among these groups, packing genes (descriptors) have the most direct and physical connection with structural formability, whereas the bonding and optoelectronic-response groups act as complementary chemically meaningful variables for describing stability and functional trends across the compositional space. Starting from a broader descriptor pool, redundant were removed through correlation analysis and feature-pruning criteria. According to Figure 5, descriptors showing only very weak correlation with the target phenotypes ($| r | < 0.10$ with either $E_g$ or the stability label derived from $E_{\mathrm{hull}}$) were excluded from the retained set. The correlation and pruning analysis show that B1-X and B3-X bond energies, together with the $B$-site ionic radius, remain particularly influential for band-gap variation, while the final retained descriptor set still preserves the coupled information needed for stability learning (see Table 3).

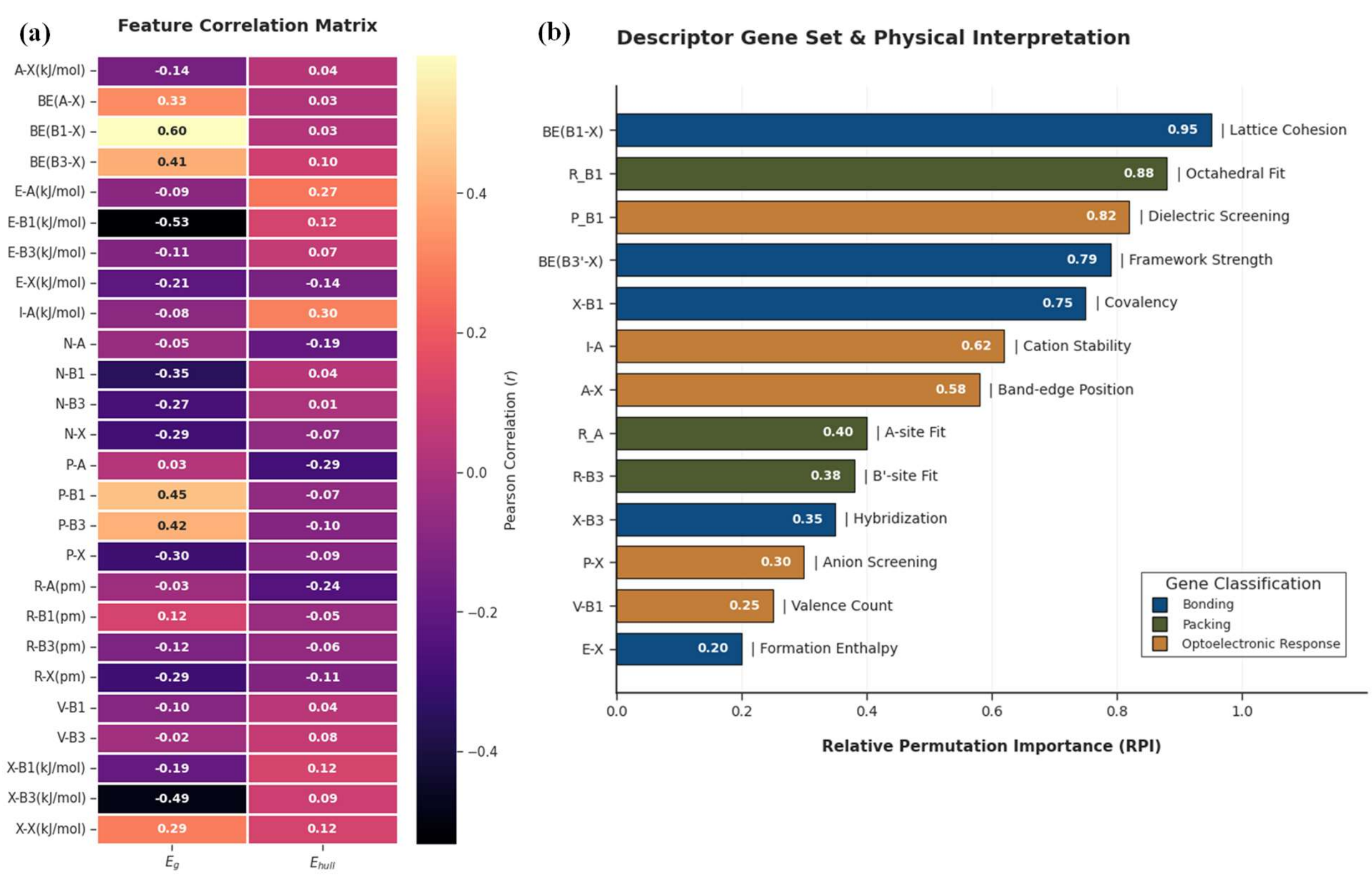

*Figure 5. Construction of the stability genome and feature selection process. (a) Correlation matrix of the descriptors used to identify highly related variables. (b) Final descriptor set with low collinearity selected for machine-learning model training.*

*Table 3. Stability-genome descriptor groups used in this work. Physically interpretable, site-resolved descriptors used for machine-learning prediction of thermodynamic stability ($E_{hull}$-derived label) and electronic band gap ($E_g$). Coordination numbers used for ionic radii were $A = 12$, $B/B' = 6$, and $X = 6$.*

| Functional role | Gene cluster | Descriptor(s) | Site | Units | Physical role |
|---|---|---|---|---|---|
| Structural existence | Packing (formability) | Ionic radii R-A, R-B1, R-B3, R-X | A, B, B′, X | pm | Determines ionic size matching, lattice distortion, and fitting of the $BX_6$ octahedral framework |

| | | Tolerance factors t, μ, τ | A/B/X | – | Indicates geometric compatibility for forming the perovskite structure |
|---|---|---|---|---|---|
| Thermo-dynamic stability | Bonding (framework cohesion) | Bond dissociation energies A-X, B1-X, B3-X | A–X, B–X, B′–X | kJ $mol^{-1}$ | Describes the strength of chemical bonding and stability of the octahedral network |
| | | Electronegativity X-B1, X-B3, X-X | B, B′, X | – | Controls bond polarity, covalency, and metal–halide interaction |
| | | Atom Formation enthalpy E-A, E-B1, E-B3, E-X | A, B, B′, X | kJ $mol^{-1}$ | Shows elemental reactivity and contribution to compound stability |
| Opto-electronic Function | Opto-electronic response | First Ionization energy I-A | A | kJ $mol^{-1}$ | Indicates cation ionization and charge-transfer tendency |
| | | Electron affinity A-X | X | kJ $mol^{-1}$ | Describes anion reducibility and charge accommodation |
| | | Polarizability P-A, P-B1, P-B3, P-X | A, B, B′, X | a.u. | Related to dielectric constant and lattice softness |
| | | Valence electrons V-A, V-B1, V-B3, V-X | A, B, B′, X | – | Orbital filling; band-edge character |
| | | Atomic number N-A, N-B1, N-B3 | A, B, B′ | – | Periodic-trend and relativistic effects |

Table 4 summarizes the main role of each descriptor group in the screening workflow. It helps to understand how the three groups contribute differently to formability, stability, and optoelectronic behavior.

*Table 4. Summary of the descriptor groups used in the genome guided framework.*

| Descriptor group | Main descriptors included | Primary role in the workflow | Main relevance in this study | Interpretation strength |
|---|---|---|---|---|
| Packing | Ionic radii, $t$, $\mu$, $\tau$ | Structural existence and geometric feasibility | Defines whether a composition is geometrically compatible with the double-perovskite framework | Strongest and most directly supported |

| Bonding | Bond dissociation energies, electronegativity differences, atomic formation enthalpies | Thermodynamic stability and framework cohesion | Captures chemical interactions relevant to bond strength, polarity, and stability trends | Strong, but best interpreted as chemically meaningful rather than fully causal |
|---|---|---|---|---|
| Opto-electronic response | Ionization energy, electron affinity, polarizability, valence electron count, atomic number | Optoelectronic screening and band-edge-related behavior | Gives information associated with dielectric response, charge redistribution, periodic trends, and electronic character | Supportive and complementary |

### 3.3 Evolutionary-optimized ML surrogates: stability inference with high stable-class recall and band-gap targeting

After constructing the descriptor genome, the next step was to learn how this genome is translated into the DFT-derived phenotypes. In the present framework, the machine-learning models are not treated only as ordinary property predictors. Instead, they are used as genome-to-phenotype mapping functions, through which compact descriptor representations are connected to thermodynamic stability and electronic response. In this way, machine learning acts as the quantitative bridge between the inverse-design object, namely the descriptor genome, and the phenotype space that later guides composition realization.

Two supervised phenotypes were considered in this study (Figure 6,7). The first phenotype was the thermodynamic stability label derived from $E_{\text{hull}}$, where compounds with $E_{\text{hull}} \leq 25$meV $atom^{-1}$ were treated as stable. The second phenotype was the scalar GGA-PBE band gap $(E_g)$, which was used for the initial optoelectronic down-selection. Therefore, the learning problem was defined as two related but different mappings: one from descriptor genomes to DFT-level thermodynamic stability, and another from descriptor genomes to DFT-level band-gap behavior. This separation is important because structural and chemical admissibility do not always determine electronic behavior in the same way. A compact genome can therefore affect more than one phenotype, but often through different descriptor families and with different relative importance.

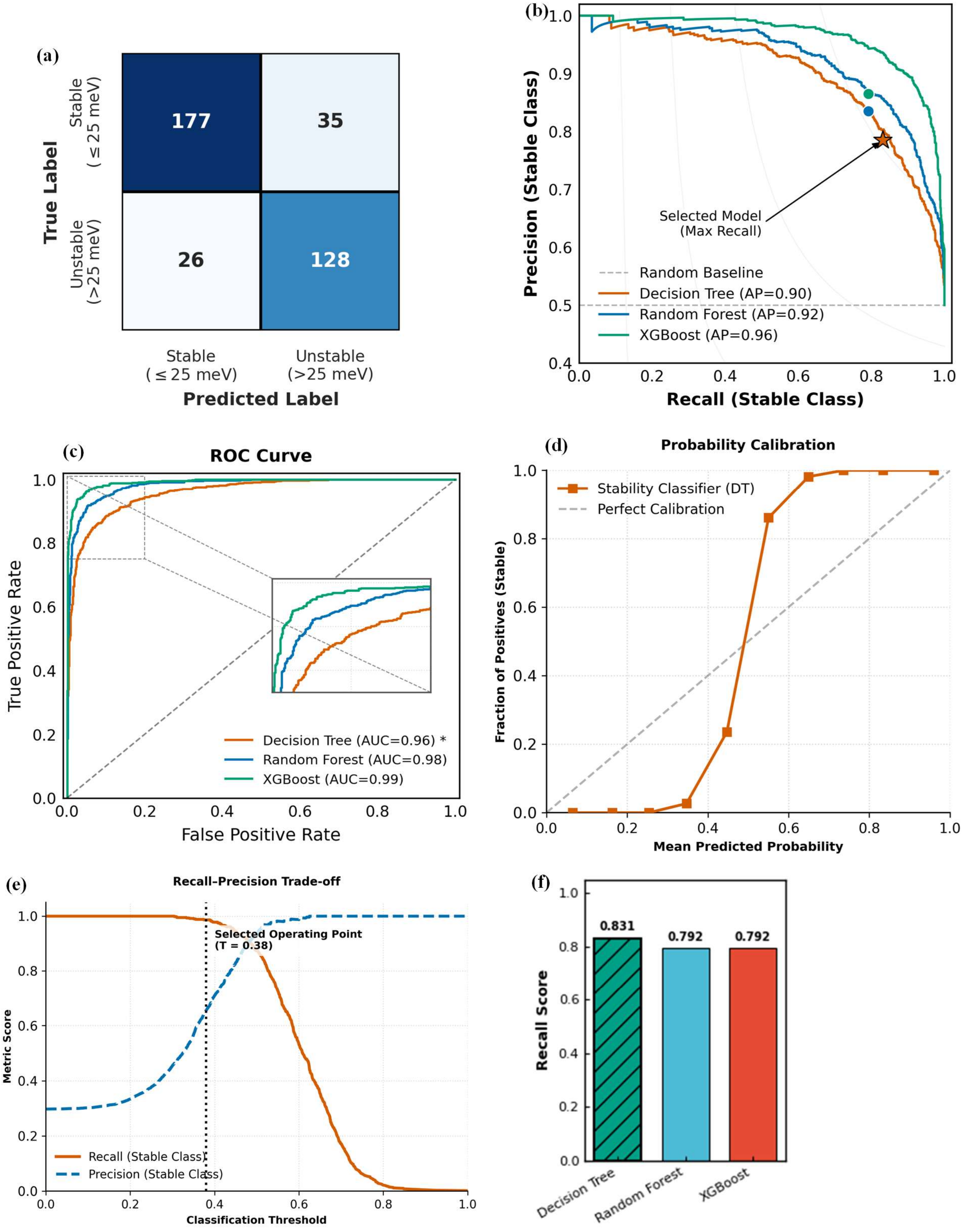

*Figure 6. Performance of the recall-optimized stability classifier. (a) The confusion matrix, (b) precision–recall curve, (c) receiver operating characteristic (ROC) curve, (d) calibration behavior, (e) operating point selected and (f) the recall score for high stable-class recalls during inverse-design screening.*

For the thermodynamic-stability mapping, Decision Tree (DT), Random Forest (RF), and XGBoost classifiers were evaluated. For the genome-to-band-gap mapping, Random Forest regression, support vector regression with a radial basis function kernel, and XGBoost regression were examined. The final model selection was not based only on overall predictive performance, but also on the role each model plays in the genome guided-design workflow. For stability classification, high recall of the stable class was especially important because false-negative predictions would remove potentially realizable compounds too early from the realization space. For the band-gap phenotype, the main objective was to preserve reliable regression performance across the screened compositional domain.

The stability results in Table 5 clearly show this difference in purpose. Although the Evolutionary Algorithm (EA)-optimized XGBoost classifier reached the highest overall accuracy (0.860), the EA-optimized Decision Tree model was selected as the primary screening model because it produced the highest stable-class recall (0.831), compared with 0.792 for both RF and XGBoost. This choice is important in the context of inverse screening. A false positive only causes additional downstream DFT cost, whereas a false negative means that a potentially stable compound is removed before realization and validation. For this reason, the genome-to-stability mapping was intentionally recall-prioritized. The receiver operating characteristic behavior also indicates strong separability for all optimized classifiers, with Arear Under Curve (AUC) values above 0.90, confirming that the descriptor genome retains strong information about thermodynamic accessibility.

*Table 5. Test-set performance of the evolutionary-optimized stability classifiers.*

| Model (Classifier) | Accuracy | Precision (stable) | Recall (stable) |
|---|---|---|---|
| RF (EA) | 0.846 | 0.835 | 0.792 |
| XGBoost (EA) | 0.860 | 0.865 | 0.792 |
| DT (EA) | 0.833 | 0.785 | 0.831 |

*Note: the positive class "Stable" is defined as $E_{hull} \leq 25$ meV atom$^{-1}$.*

The probability calibration and recall-precision analysis further support the usefulness of the selected stability model. The calibration curve of the Decision Tree classifier shows that predicted stability probabilities become meaningful mainly in the intermediate-to-high probability region, which is acceptable for the present workflow because the model is not used as a final thermodynamic authority, but as a realization-stage filter before DFT phenotype closure. The recall-precision trade-off also shows that the selected operating region preserves very high recall while keeping reasonable precision, which is exactly the desired behavior when the main objective is to avoid losing potentially stable compositions too early.

The genome-to-band-gap mapping shows similarly strong performance. Among all tested models, the Evolutionary Algorithm (EA)-optimized XGBoost model performs the best. It achieves an $R^2$ value of about 0.93. The RMSE is about 0.51 eV, and the MSE is about 0.26 eV² (Table 6 and Figure 7). The predicted band gaps and the DFT bandgap are very close. Most of the data points lie near the identity line. This trend is consistent across the full 0–8 eV range. This indicates that the model does not show strong bias. It performs well for both low and high band-gap values (Figure 7). The residual distribution is centered close to zero, and the error-versus-$E_g$ analysis shows that the model remains reasonably stable across the full target range. These results indicate that the descriptor genome is sufficiently informative to capture band-gap variation across chemically diverse lead-free double perovskites with near-DFT fidelity at the screening level.

*Table 6. Test-set performance of the evolutionary-optimized band-gap regressors.*

| Model (Regressor) | $R^2$ | RMSE (eV) | MSE (eV²) |
|---|---|---|---|
| SVR (RBF, EA-tuned) | 0.9244 | 0.5414 | 0.2931 |
| Random Forest (EA-tuned) | 0.9241 | 0.5426 | 0.2944 |
| XGBRegressor (EA-tuned) | 0.9317 | 0.5144 | 0.2646 |

The evolutionary convergence behavior shows that the optimized band-gap surrogate is not obtained by chance. The validation MSE drops quickly in the early generations. After that, it slowly levels off to a stable, low value. This indicates that the genetic algorithm effectively guides the search. It moves the population toward better regions of the hyperparameter space and improves model fitness in a consistent way. This behavior gives further confidence that the final XGBoost regressor is a robust genome-to-band-gap mapping rather than a locally unstable fit.

An important aspect of this step is that the DFT-labeled dataset and the enumerated realization space were generated independently, even though both belong to the same $A_2BB'X_6$ halide double-perovskite family. This means that the learned genome-to-phenotype relationships were not simply memorized from the realization space itself but were transferred from the independently assembled DFT dataset to the broader enumerated library. As a result, the surrogate models function as learned phenotype maps that guide which regions of the realization space are most likely to express the desired descriptor patterns.

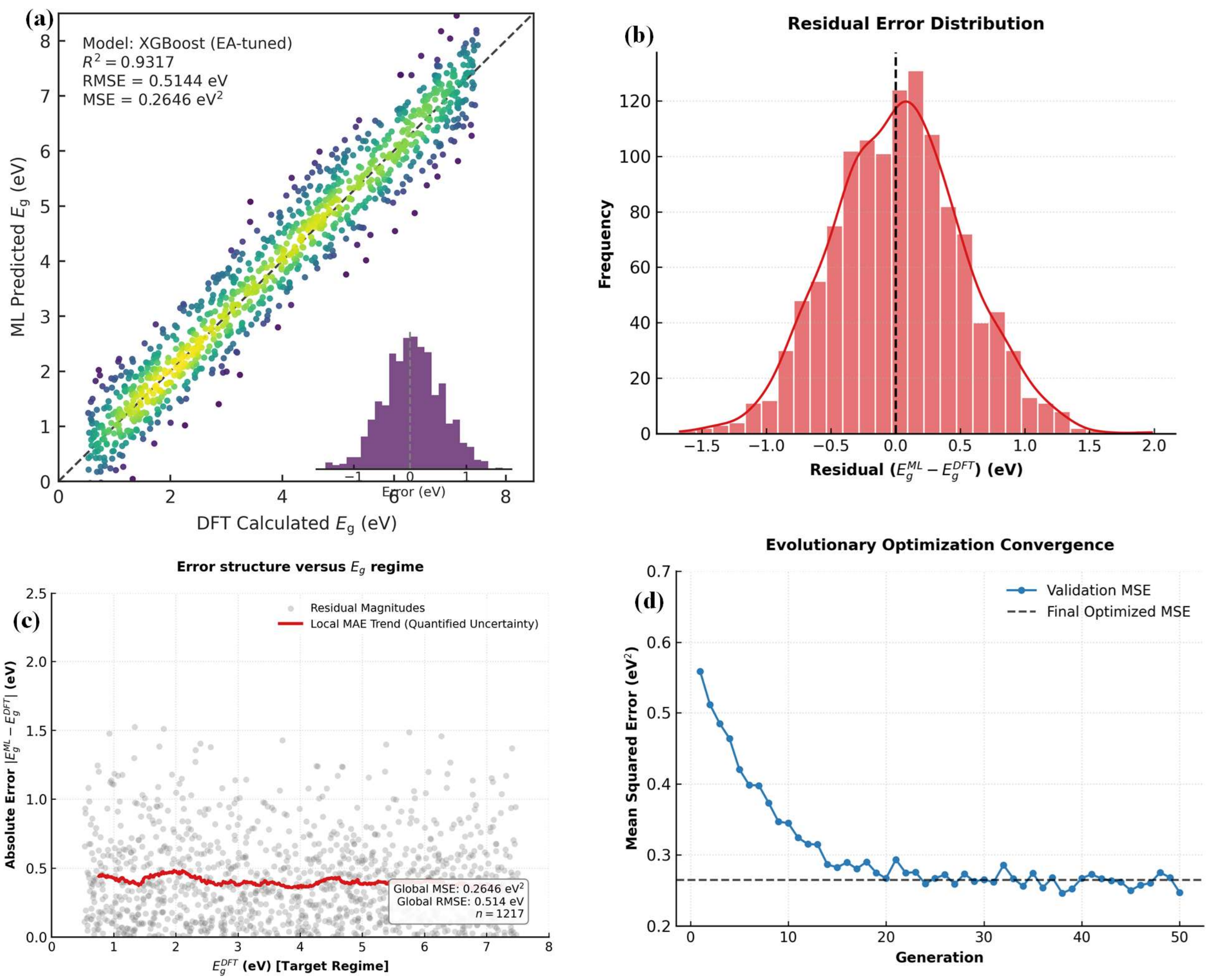

*Figure 7. Performance of the evolutionary-optimized band-gap regressor. The figure summarizes (a) parity between predicted and DFT-calculated E_g, (b) residual statistics, (c) error distribution across the band-gap range, and (d) evolutionary convergence during hyperparameter optimization.*

The interpretability analysis further supports the physical meaning of these mappings. From Figure 5 and 8, permutation importance and partial dependence analysis show that the optimized models do not rely on arbitrary statistical artifacts but recover chemically sensible trends. In the XGBoost regressor, B1-X bond energy and *B*-site polarizability (P-B1) emerged as dominant predictors, followed by *B*-site electronegativity, *B*-site ionic radius, and B1-X, B3-X bond dissociation energy. The partial dependence plots show physically consistent monotonic trends, in which stronger B1-X bonding and higher *B*-site electronegativity are associated with larger predicted band gaps. The two-feature interaction landscape further indicates that the band gap is not controlled by a single descriptor independently, but by coupled variation in bonding strength and polarizability. This result is fully consistent with the descriptor-genome logic established in Section 3.2, especially the importance of bonding and *B*-site-centered response terms.

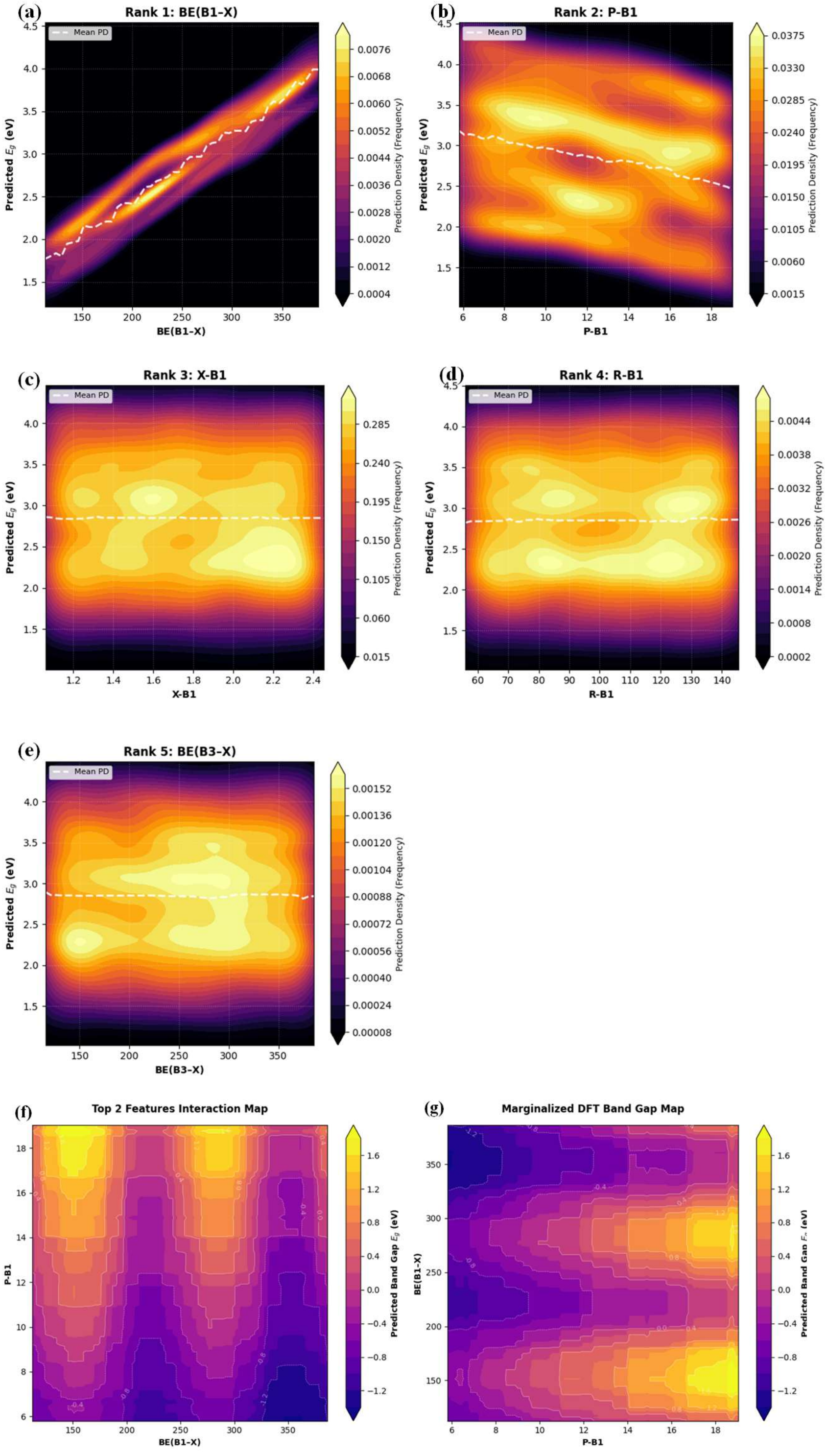

*Figure 8. Partial dependence analysis of the top five ranked descriptors controlling the predicted band gap. The one-dimensional PDPs show how each feature affects the predicted band gap when considered separately. The features are ranked by importance. BE(B1–X) is the most important feature. P–B1 comes next. X–B1, R–B1, and BE(B′–X) follow in order. Figure (f) shows the interaction between the two most important features, BE(B1–X) and P–B1. It shows how their combined effect changes the band gap prediction. Figure (g) compares the model trends with DFT results. It plots the calculated band gaps in the P–B1 and BE(B1–X) space. The color scale shows the band-gap values in eV. The contour lines help to show how the values change across the map.*

Within the present framework, the value of these surrogate models lies not only in prediction speed, but also in the fact that they make descriptor-level design rules operational. Once the genome-to-phenotype mappings are established, they can be used to move from compact descriptor combinations to specific composition candidates that are likely to realize those combinations. Therefore, machine learning is used here not as a black-box ranking tool, but as the computational engine that connects inverse design in descriptor space to phenotype-guided realization in composition space.

### 3.4 Mechanistic genome decoding: separating feasibility, stability, and optoelectronic control

After establishing the genome-to-phenotype mappings, the next step is to interpret how the descriptor genome governs the learned phenotypes in physical terms. In this framework, the genome is not only a compact representation of composition, but a structured control space that enables translation of model behavior into mechanistic design rules. A useful inverse-design strategy must therefore move beyond prediction and identify which chemical variables should be tuned to steer the phenotype in a targeted direction.

Figure 9 provides this mechanistic decoding at two complementary levels. First, SHAP summaries in Figure 9(a)-(b) quantify how individual genes (descriptors) contribute to the predicted stability and band-gap phenotypes. Second, the response curves in Figure 9(c)–(e) translate these contributions into physically interpretable control rules governing structural feasibility, bonding strength, and optoelectronic tuning. Together, these analyses show that different descriptor families act as distinct but coupled control variables within the genome.

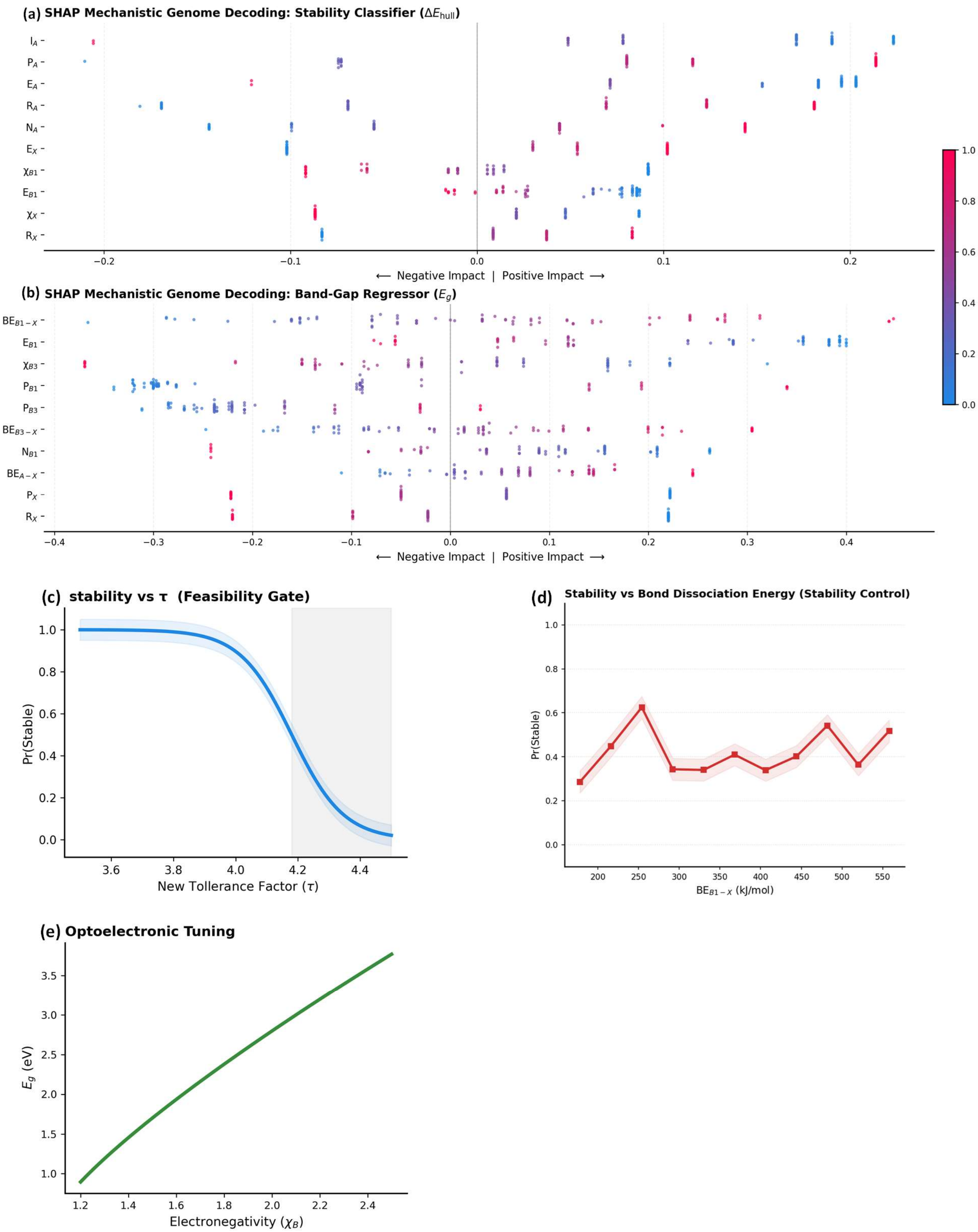

*Figure 9. Mechanistic genome decoding from feature attribution and response analysis. (a) SHAP summary for thermodynamic stability. (b) SHAP summary for band gap. (c) Stability probability as a function of τ. (d) Stability response versus B–X bond dissociation energy. (e) Band-gap response versus B-site electronegativity.*

For thermodynamic stability, the SHAP analysis (Figure 9(a)) indicates that the stable class emerges from a combination of *A*-site response descriptors, formation-enthalpy-related variables, ionic-size metrics, and halide chemistry. This shows that stability depends on multiple linked factors. These include geometric fit, bonding strength, and elemental properties. No single descriptor is enough to explain it. In contrast, the band-gap behavior depends on a more focused set of key descriptors (Figure 9(b)).

The dominant contributors are *B–X* and *B′–X* bond energies, *B*-site polarizability, electronegativity, and ionic radius, highlighting the central role of the octahedral environment. This is physically consistent with the $BX_6$ and $B'X_6$ frameworks forming the electronic backbone of double perovskites, where band edges are primarily defined by metal–halide hybridization. Thus, while structural feasibility remains necessary, band-gap tuning is controled more directly by bonding and electronic response within the octahedral sublattice.

The response curves in Figures 9(c)–(e) further clarifies this hierarchy. The stability probability as a function of $\tau$ (Figure 9(c)) shows a sharp decline beyond the formable region, confirming that packing descriptors act as a feasibility gate. Structural compatibility is therefore a strict prerequisite; once a composition lies outside this geometric window, further optimization becomes ineffective. Within the formable region, bonding descriptors become the primary control variables. Figure 9(d) shows that stronger *B–X* bond dissociation energy generally increases the probability of stability, reflecting enhanced framework cohesion in the corner-sharing octahedral network.

Finally, Figure 9(e) shows that optoelectronic properties can be systematically tuned using response genes (descriptors). The band gap increases with *B*-site electronegativity. It indicates that electronic identity and bonding characteristics at the octahedral site can be effective levers to control band-edge behavior. Importantly, this trend represents a coupled response between bonding strength, polarizability, and electronegativity, rather than a single-variable effect.

The key outcome is that descriptor-level interpretability is converted into phenotype-linked design logic. The genome not only predicts viable candidates but explains why specific combinations of structural fit, bonding chemistry, and electronic response lead to realizable and functional double perovskites. This mechanistic understanding forms the basis for translating descriptor-space design rules into composition-space realization in the subsequent screening stage.

### 3.5 Genome-guided interpretable screening and composition realization

After establishing the descriptor genome and the genome-to-phenotype relationships, the next step was to translate this design logic into real $A_2BB'X_6$ compositions. In the present framework, screening is used as a realization step, not only as a ranking step. Each stage tests whether a composition can satisfy one required layer of the genome-guided criteria.

As summarized in Figure 10 and Table 6, the initial library contained 13,088 charge-balanced lead-free compositions. In Stage I, the recall-prioritized stability classifier retained 5,354 candidates predicted to belong to the stable class. In Stage II, geometry screening using $t$, $\mu$, and $\tau$ reduced the set to 733 compositions, showing that structural admissibility is one of the strongest restrictions in the chemical space. In Stage III, band-gap targeting further reduced the candidate set to 162 compositions inside the desired optoelectronic window. After stricter deployability filters, only 24 candidates remained, and targeted DFT validation finally reduced the set to 5 phase-stable semiconductor candidates. Overall, only 0.04% of the initial library survived the full screening workflow.

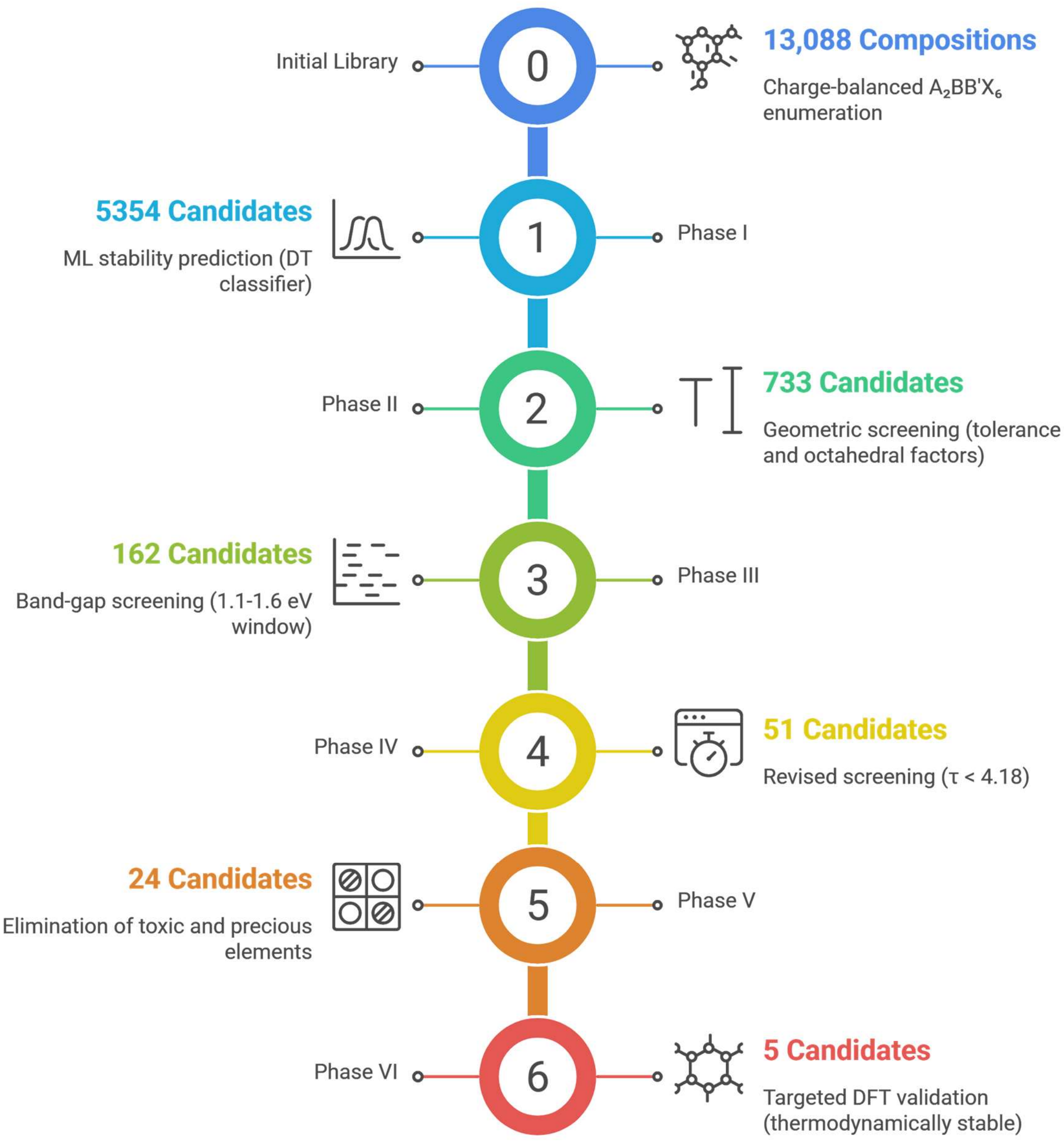

*Figure 10. Inverse-design constraint stack and candidate funnel. The panels summarize candidate reduction across screening stages, the predicted stability–band-gap landscape, and the application windows used to identify the final DFT-validation set.*

This narrow overlap is also visible in Figure 11, where the predicted stability probability is plotted against the predicted band gap. Most compositions lie outside the desirable region combining high stability probability and useful gap windows. In contrast, the DFT-confirmed compounds occupy a small and isolated area, marked here as the rare stability-function intersection. This result shows that useful candidates do not come from the entire chemical space. They come only from a small and specific region. In this region, all key conditions are satisfied together. These include structural feasibility, thermodynamic stability, and optoelectronic relevance.

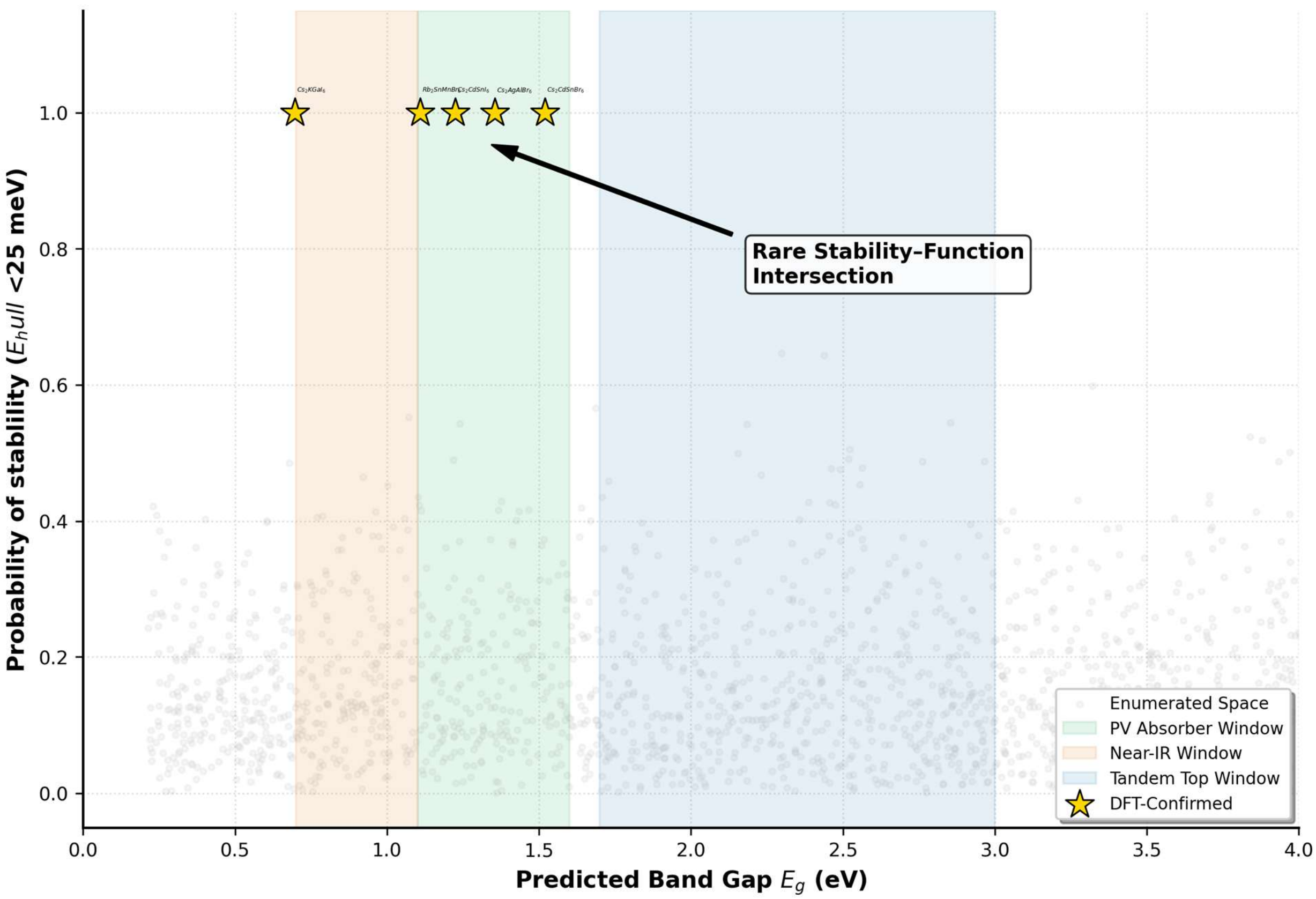

*Figure 11. Plot for the predicted stability probability vs the predicted band gap.*

Summary of the staged workflow used to translate descriptor-genome rules into composition-space candidates shown in Table 7. The progressive reduction highlights the narrow overlap between structural feasibility, thermodynamic stability, target band gap, and deployability in the lead-free $A_2BB'X_6$ chemical space.

Importantly, the number of candidates decreased strongly at each screening stage. This shows that only a small part of the design space is suitable. Many materials do not satisfy all the required properties. The materials must have structural stability, thermodynamic feasibility, and good optoelectronic properties at the same time. These requirements create a complex selection region. Therefore, multi-objective screening methods are more effective than single-property optimization. Interpretable screening approaches are also important.

*Table 7. Multi-stage screening criteria and candidate reduction statistics.*

| Screening stage | Candidates remaining | Fraction of initial (%) | Main criterion | Notes |
|---|---|---|---|---|
| Initial library | 13,088 | 100.00 | Charge-balanced $A_2BB'X_6$ enumeration | Lead-free compositions only |

| Stage I | 5,354 | 40.91 | ML stability classifier | Predicted stable compositions |
|---|---|---|---|---|
| Stage II | 733 | 5.60 | Geometry screening | $0.80 \le t \le 1.10$; $0.414 \le \mu \le 0.732$; $\tau < 4.18$ |
| Stage III | 162 | 1.24 | Band-gap targeting | Application-oriented Eg window |
| Stage IV | 24 | 0.18 | Deployability filters | Tighter chemistry and admissibility filters |
| Stage V | 5 | 0.04 | Targeted DFT validation | Phase-stable semiconductor candidates |

## 3.6 DFT phenotype closure: thermodynamic stability, structure, and electronic origin

DFT closure is the decisive test of genome guided screening and design rule: it confirms that ML-selected candidates are not only plausible but exhibit the structural and electronic characteristics required for functional optoelectronics.

### *3.6.1 DFT-validated phase-stable semiconductors*

Targeted DFT validation identifies five phase-stable semiconductor candidates within the screened chemical space: $Rb_2SnMnBr_6$, $Cs_2CdSnBr_6$, $Cs_2CdSnI_6$, $Cs_2KGaI_6$, and $Cs_2AgAlBr_6$ (see Table 8). These compounds are located on or very close to the convex hull within the selected phase reference. They also maintain the ordered double-perovskite structure after relaxation. $Cs_2CdSnI_6$ keeps its ideal cubic symmetry. The other compounds show small symmetry changes after relaxation. However, they still maintain corner-sharing octahedra and clear B/B′ ordering.

*Table 8. DFT-validated realizations of the descriptor-genome design rules. Key structural and electronic metrics of the five phase-stable $A_2BB'X_6$compounds that realize the inverse-designed descriptor genomes, including PBE band gap, geometric descriptors, energy above hull, space group, and lattice parameter.*

| Compound | Bandgap, $E_g$ (eV) | t | μ | τ | Space group | Lattice parameter a (Å) |
|---|---|---|---|---|---|---|
| $Rb_2SnMnBr_6$ | 1.237 | 0.853 | 0.472 | 4.084 | P1 | 12.4 |
| $Cs2CdSnBr_6$ | 1.794 | 0.849 | 0.543 | 4.175 | P1 | 11.488 |
| $Cs2CdSnI_6$ | 1.132 | 0.838 | 0.484 | 4.179 | Fm-3m | 8.238 |
| Cs2KGaI6 | 0.696 | 0.837 | 0.486 | 4.131 | P 1 | 12.453 |
| $Cs_2AgAlBr_6$ | 1.096 | 0.899 | 0.457 | 4.131 | P1 | 11.00 |

*These five materials were chosen for detailed structural, electronic, and optical studies in the following sections.*

### *3.6.2 Structural assignability: Crystal structure and XRD fingerprints*

Figure 12 shows the optimized crystal structures and simulated powder X-ray diffraction (XRD) patterns of the five selected double perovskites. The relaxed structures keep the ordered $A_2BB'X_6$ framework. The diffraction patterns indicate good crystallinity and clear long-range B/B′ ordering. All compounds show sharp diffraction peaks. No clear impurity peaks are observed. A weak low-angle (111) superlattice peak is also present. This peak is important because it confirms rock-salt ordering between the two octahedral sublattices. The peak positions change across the series. This mainly depends on lattice size and halide type. The iodide-based compounds show peaks at lower angles because their lattices are larger. The bromide-based systems show peaks at higher angles due to their smaller lattice size. The relative peak intensities also vary. This comes from differences in atomic scattering and cation contrast at the B and B′ sites. Overall, the simulated structures and XRD patterns confirm the structural assignment of the selected materials. They also support their likelihood of being experimentally realizable.

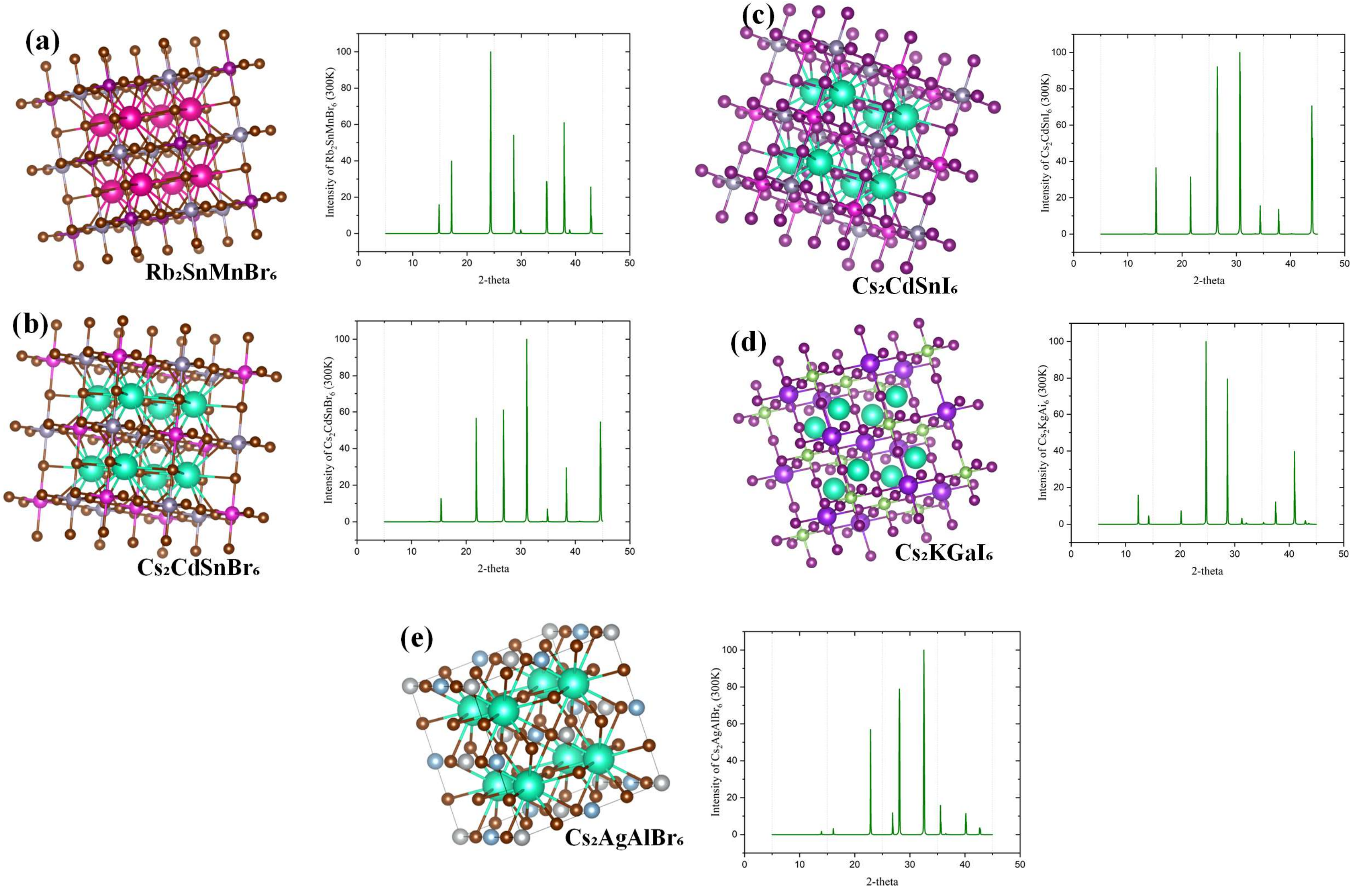

*Figure 12.Optimized crystal structures and simulated powder X-ray diffraction patterns of (a) $Rb_2SnMnBr_6$, (b) $Cs_2CdSnBr_6$, (c) $Cs_2CdSnI_6$, (d) $Cs_2KGaI_6$, and (e) $Cs_2AgAlBr_6$. All compounds show sharp diffraction peaks. It indicate good crystallinity. A weak (111) superlattice peak is also observed. This confirms the ordered double-perovskite structure and preserved B/B' ordering in all cases.*

Overall, the sharp diffraction peaks and the absence of impurity peaks show that all five compounds have stable and ordered double-perovskite structures, as summarized in Table 8. The weak (111) superlattice peak also supports the presence of B/B′ ordering. This ordered structure is useful for good optoelectronic performance. In contrast, disorder at the B-site can increase carrier scattering and create trap states.

*Table 8. Structural assignability and XRD fingerprints of the DFT-realized compounds. Major reflections, superlattice features, and ordering signatures extracted from the simulated powder XRD patterns.*

| **Compound** | **Main reflection (hkl)** | **Major peaks (2θ, °)** | **(111) peak (2θ, °)** | **Phase purity/ ordering notes** | **Structural meaning** | **Intensity** |
|---|---|---|---|---|---|---|
| $Rb_2SnMnBr_6$ | (220) | 16.2, 23.5, 28.7, 39.6 | 13.4 | No impurity peaks are seen. A clear superlattice peak is present. | Sn and Mn are ordered in a rock-salt structure. | High |
| $Cs_2CdSnBr_6$ | (220)/ (222) | 17.0, 24.8, 30.2, 44.6 | 13.8 | The pattern is clean with sharp peaks. | Cd and Sn give similar scattering, which leads to balanced peak intensities. | High |
| $Cs_2CdSnI_6$ | (222) | 16.0, 22.8, 28.9, 45.8 | 12.6 | The compound shows a pure cubic phase. | Larger iodine ions expand the lattice and shift peaks to lower angles. | High |
| $Cs_2KGaI_6$ | (200) | 13.5, 24.2, 29.6, 40.5 | 11.8 | A clear superlattice peak confirms ordering. | K and Ga are well ordered in the structure. | High |
| $Cs_2AgAlBr_6$ | (220) | 21.4, 27.9, 31.5, 38.7 | 13.2 | No extra peaks are seen. The structure is well ordered. | $AgBr_6$ and $AlBr_6$ octahedra alternate regularly and stabilize the cubic phase. | High |

### *3.6.3* Electronic origin: band structures and PDOS reveal band-edge chemistry

The final part of this section is to examine how the descriptor genome is expressed in the electronic structure (see Figure 13). The band structures and projected density of states (PDOS) provide the orbital-level explanation for how the descriptor genome is expressed in the shortlisted compounds. All five materials show indirect or weakly indirect semiconducting behavior, with the valence-band maximum mainly dominated by halide p states and the conduction-band minimum mainly formed by cation s/p states. This common band-edge pattern is important because it explains why metal-halide hybridization repeatedly appeared as a key bonding feature in the machine-learning interpretation. In this way, the descriptor-genome logic is not only statistical, but is also reflected in the electronic structure of the realized compounds.

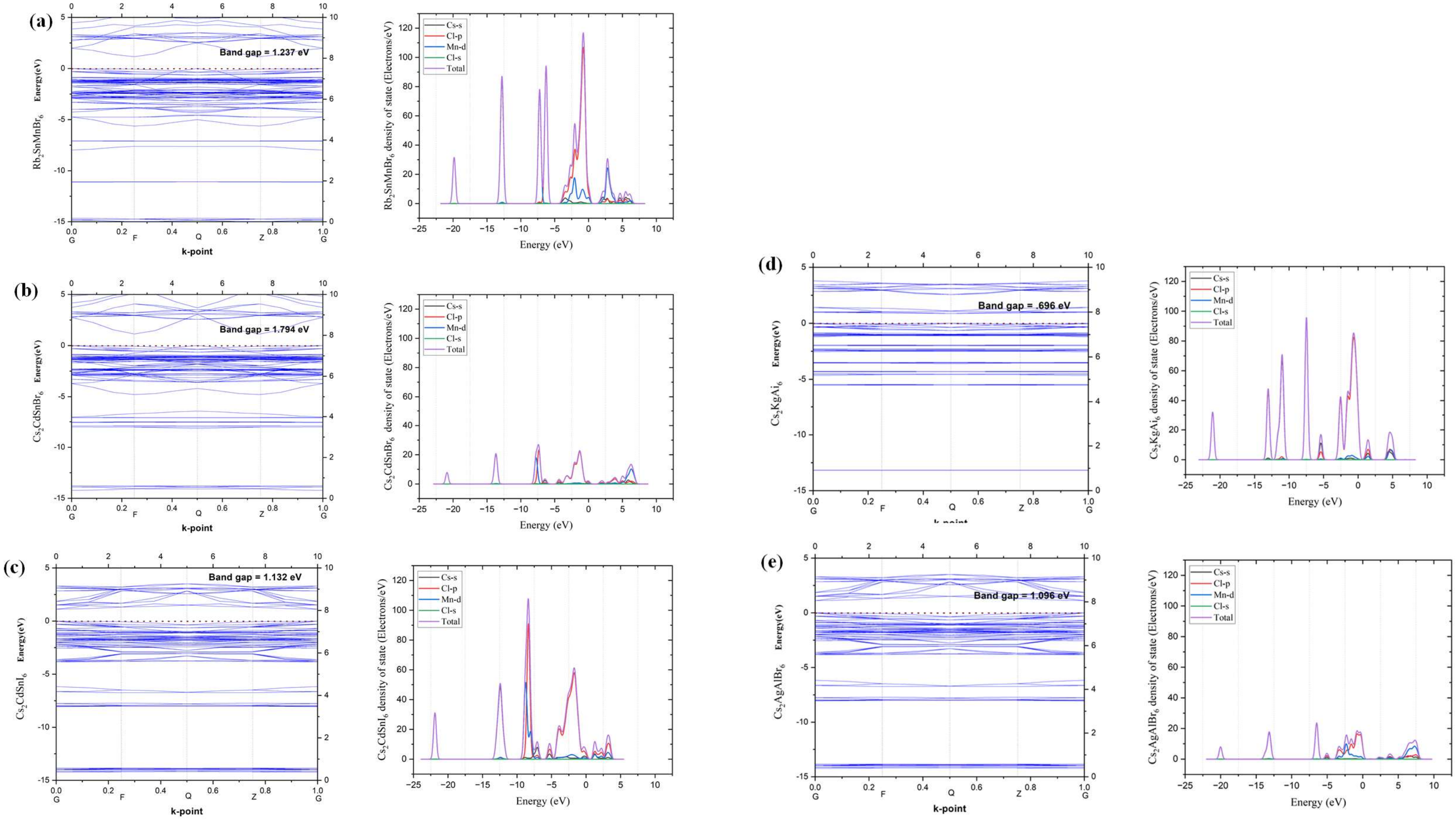

*Figure 13. Electronic band structures and projected density of states (PDOS) of the DFT-validated double perovskites. Panels (a–e) show $Rb_2SnMnBr_6$, $Cs_2CdSnBr_6$, $Cs_2CdSnI_6$, $Cs_2KGaI_6$, and $Cs_2AgAlBr_6$, respectively. The band structure and PDOS results show that all compounds have indirect band gaps. The valence bands are mainly formed by halide p states. The conduction band edges are mainly formed by cation states. The Mn-containing compound shows stronger localization of the valence states compared to the others.*

At the same time, the five compounds remain electronically different in a chemically meaningful way. $Rb_2SnMnBr_6$ is distinguished by stronger Mn 3d contribution near the valence-band edge, which produces more localized valence states, heavier holes, and stronger transport asymmetry. In contrast, $Cs_2CdSnBr_6$ shows the clearest separation between s- and p-orbital contributions at the band edges. It has more dispersive valence and conduction bands. It also has the lightest carrier effective masses among the studied compounds. When Br is replaced by I in $Cs_2CdSnI_6$, the higher-energy I 5p states raise the valence-band edge. This also increases lattice polarizability. However, the band dispersion becomes narrower. As a result, the effective masses increase. $Cs_2KGaI_6$ also shows an iodide-dominated valence-band edge. However, its conduction band is flatter due to Ga states. This leads to weaker charge transport compared to the Cd/Sn-based compounds. For $Cs_2AgAlBr_6$, the conduction band remains more dispersive than the valence band, resulting in electron-favored transport within a relatively more rigid bromide framework.

These trends define a direct genome-to-band-edge pathway: descriptor tuning affects orbital hybridization, orbital hybridization shapes band dispersion, and band dispersion then influences carrier asymmetry and transport behavior (Table 9). The effective-mass trends follow the same logic and provide a transport-level readout of the descriptor chemistry. Therefore, the descriptor genome does not stop at structural feasibility or scalar property prediction. It also extends to the orbital-level origin of the phenotype. This is an important physical validation of the genome-guided framework, because it shows that the learned descriptor rules are realized not only in phase stability and structural assignability, but also in the underlying electronic structure of real lead-free $A_2BB'X_6$ compounds.

*Table 9. Carrier effective masses and transport descriptors of DFT-validated double perovskites.*

| Compound | m_e* ($m_0$) | m_h* ($m_0$) | Electron mobility trend | Hole mobility trend | Band-edge character |
|---|---|---|---|---|---|
| $Rb_2SnMnBr_6$ | 0.18 | 2.10 | High (light electrons) | Very low (strongly localized holes) | Sn 5s/p-derived CBM; Mn 3d–Br 4p VBM; pronounced charge-transfer character and strong carrier asymmetry |
| $Cs_2CdSnBr_6$ | 0.087 | 0.197 | Very high (ultralight electrons) | Moderate–high | Sn 5s/p CBM; Br 4p VBM; highly dispersive bands with near-ambipolar transport characteristics |

| $Cs_2CdSnI_6$ | 0.73 | 1.06 | Moderate | Low–moderate | Sn 5s/p CBM; I 5p–Sn 5s VBM; enhanced spin–orbit coupling and lattice expansion lead to heavier carriers |
| --- | --- | --- | --- | --- | --- |
| $Cs_2KGaI_6$ | 0.95 | 1.24 | Moderate | Low–moderate | Ga s/p CBM; I 5p VBM; relatively flat band edges yielding reduced carrier mobility |
| $Cs_2AgAlBr_6$ | 0.17 | 1.06 | High | Low–moderate | Ag/Al s/p CBM; Br 4p VBM; asymmetric transport with electron-dominated conduction |

3.7 Optical phenotype suite: ε(ω), α(ω), n(ω), R(ω), loss function, and optical conductivity

A robust inverse-design framework requires validation beyond band-gap prediction. We therefore evaluate a comprehensive optical phenotype suite derived consistently from the complex dielectric function ε(ω). In this framework, α(ω) quantifies light harvesting, n(ω) and R(ω) define optical impedance and interface constraints, the loss function L(ω) fingerprints excitation regimes, and σ(ω) captures transition strength.

*3.7.1 Dielectric function*

Within the genome guided framework, the dielectric response represents a direct phenotype of packing, bonding, and polarization descriptors. The complex dielectric function $\varepsilon(\omega) = \varepsilon_1(\omega) + i\varepsilon_2(\omega)$ links lattice polarizability, metal–halide hybridization, and band-edge composition to screening behavior (see Figure 14). All shortlisted compounds exhibit pronounced low-energy $\varepsilon_2$ peaks, confirming strong inter-band transitions near the optical onset. The Cd/Sn systems show the broadest and most intense $\varepsilon_2$ response, consistent with enhanced metal–halide hybridization.

The real part of the dielectric function, $\varepsilon_1(\omega)$, especially its low-energy limit $\varepsilon_0$, shows clear differences in screening strength among the compounds. $Cs_2CdSnI_6$ has the strongest screening, with $\varepsilon_0 \approx 8.16$. $Rb_2SnMnBr_6$ follows, with $\varepsilon_0 \approx 6.15$. $Cs_2CdSnBr_6$ shows moderate screening, with $\varepsilon_0 \approx 5.14$. $Cs_2KGaI_6$ and $Cs_2AgAlBr_6$ show the weakest screening, with $\varepsilon_0$ values of about 4.58 and 4.57, respectively.

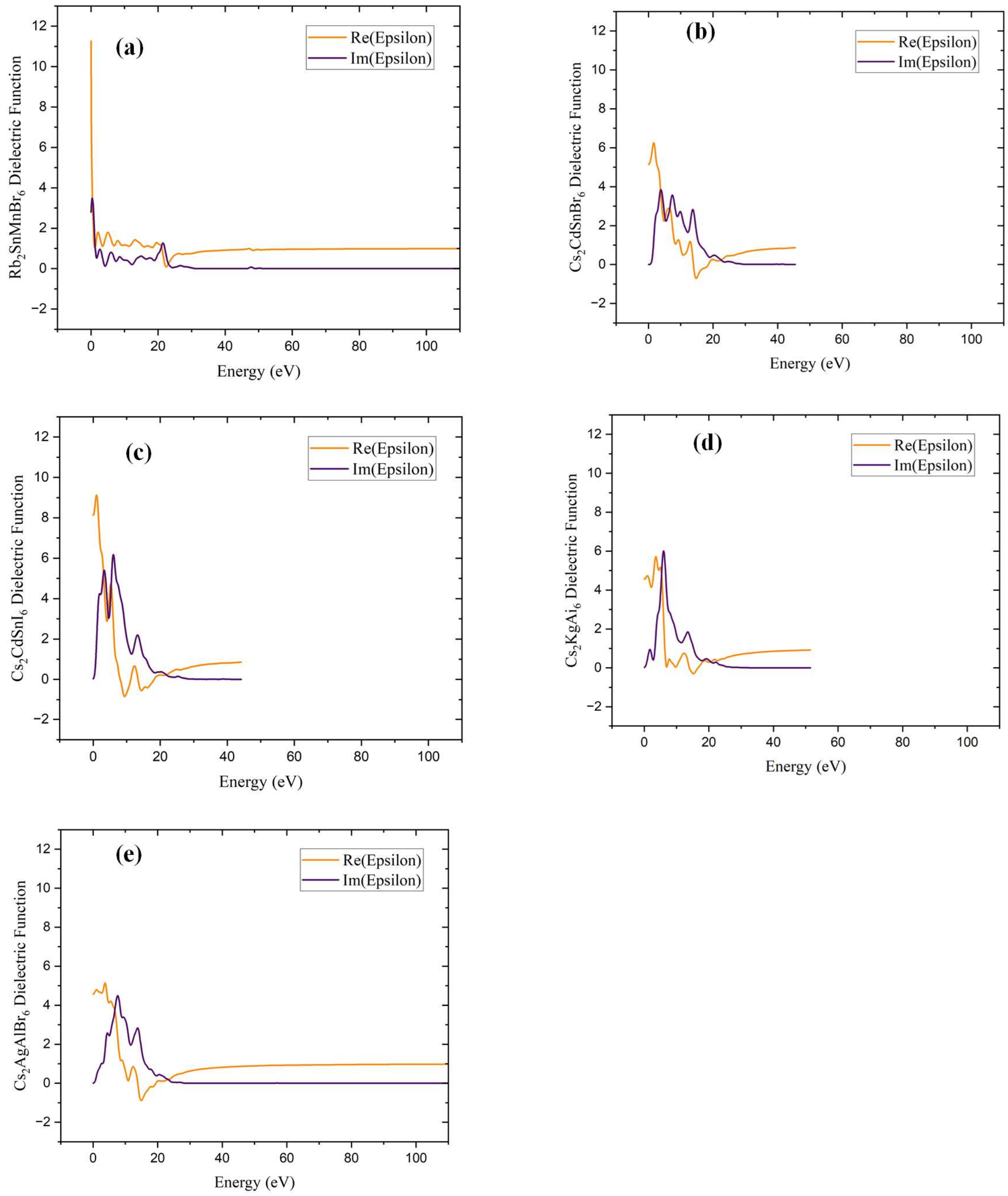

*Figure 14. Frequency-dependent complex dielectric function, $\varepsilon(\omega) = \varepsilon_1(\omega) + i\varepsilon_2(\omega)$, of the studied lead-free double perovskites: (a) $Rb_2SnMnBr_6$, (b) $Cs_2CdSnBr_6$, (c) $Cs_2CdSnI_6$, (d) $Cs_2KGaI_6$, and (e) $Cs_2AgAlBr_6$. Strong low-energy $\varepsilon_2$ peaks indicate dominant inter-band transitions, whereas the low-energy $\varepsilon_1$ baseline reflects dielectric constant strength.*

According to Table 10, the imaginary part $\varepsilon_2(\omega)$ decreases at higher photon energies. At the same time, the real part $\varepsilon_1(\omega)$ becomes almost constant. This means that most optical activity occurs in the low to mid-energy range, which is important for device applications. These results show a clear link between descriptors and properties. Opto-electronic response genes (descriptors) mainly control dielectric screening. Bonding-related descriptors mainly control the strength of inter-band transitions.

*Table 10. Dielectric function metrics relevant to screening and inter-band-transition strength.*

| Material | $\varepsilon_0$ | $\varepsilon_\infty$ | E_peak($\varepsilon_2$) (eV) | $\varepsilon_2$,max | Note |
|---|---|---|---|---|---|
| $Rb_2SnMnBr_6$ | 6.15 | 1.0 | 3–5 | 3.5 | Moderate dielectric constant and weaker inter-band transitions |
| $Cs_2CdSnBr_6$ | 5.14 | 1.0 | 6–10 | 4.0 | Enhanced absorption with improved dielectric constant |
| $Cs_2CdSnI_6$ | 8.16 | 1.0 | 5–8 | 6.0 | Strongest screening due to high polarizability of iodine |
| $Cs_2KGaI_6$ | 4.58 | 1.0 | 6–9 | 6.0 | Balanced dielectric response with moderate screening |
| $Cs_2AgAlBr_6$ | 4.57 | 1.0 | 5–10 | 4.5 | Reduced screening due to rigid lattice and lower polarizability |

### 3.7.2 Optical Absorption

The optical absorption coefficient $\alpha(\omega)$ shows how well a material can absorb light. It is directly connected to its light-harvesting ability. All selected materials show strong optical absorption. Their peak values are around $10^5$ $cm^{-1}$. This indicates that they can effectively absorb light even in thin-film form. This makes thin-film device applications realistic at the current screening stage (Table 11).

*Table 11. Dielectric constant and absorption metrics relevant to photovoltaics and photodetection.*

| Material | $\varepsilon_0$ (static) | $\varepsilon_\infty$ (high freq.) | n(0) | Peak $\alpha$ ($cm^{-1}$) | $\alpha$(Eg + 0.5 eV) ($cm^{-1}$) |
|---|---|---|---|---|---|
| $Rb_2SnMnBr_6$ | 6.15 | 1.0 | 2.48 | $2.5\times10^5$ | $1.2\times10^5$ |
| $Cs_2CdSnBr_6$ | 5.14 | 1.0 | 2.27 | $2.7\times10^5$ | $1.5\times10^5$ |
| $Cs_2CdSnI_6$ | 8.16 | 1.0 | 2.86 | $2.4\times10^5$ | $1.8\times10^5$ |
| $Cs_2KGaI_6$ | 4.58 | 1.0 | 2.14 | $2.1\times10^5$ | $1.1\times10^5$ |
| $Cs_2AgAlBr_6$ | 4.57 | 1.0 | 2.14 | $2.8\times10^5$ | $1.3\times10^5$ |

From the genome perspective, absorption is primarily governed by bonding and electronic-identity descriptors, which determine orbital overlap and transition probability. $Cs_2CdSnBr_6$ and $Cs_2CdSnI_6$ show the strongest and broadest above-onset absorption, while $Rb_2SnMnBr_6$ and $Cs_2KGaI_6$ exhibit weaker near-edge response due to stronger localization or heavier carriers. $Cs_2AgAlBr_6$ maintains strong high-energy absorption but a comparatively weaker onset. The near-edge metric α(E_g + 0.5 eV) provides a practical screening criterion. In this space, the Cd/Sn systems define the leading edge, demonstrating that optimized bonding and polarization descriptors can simultaneously promote stability and light harvesting. The spectra (Figure 15) confirm strong inter-band absorption across the low- to mid-energy region, with clear material-dependent variations in onset and intensity. Higher $\varepsilon_0$ and α(E_g + 0.5 eV) values for $Cs_2CdSnI_6$ and $Cs_2CdSnBr_6$ indicate improved dielectric screening and stronger near-edge light harvesting, which are both favorable for exciton dissociation and photocurrent generation.

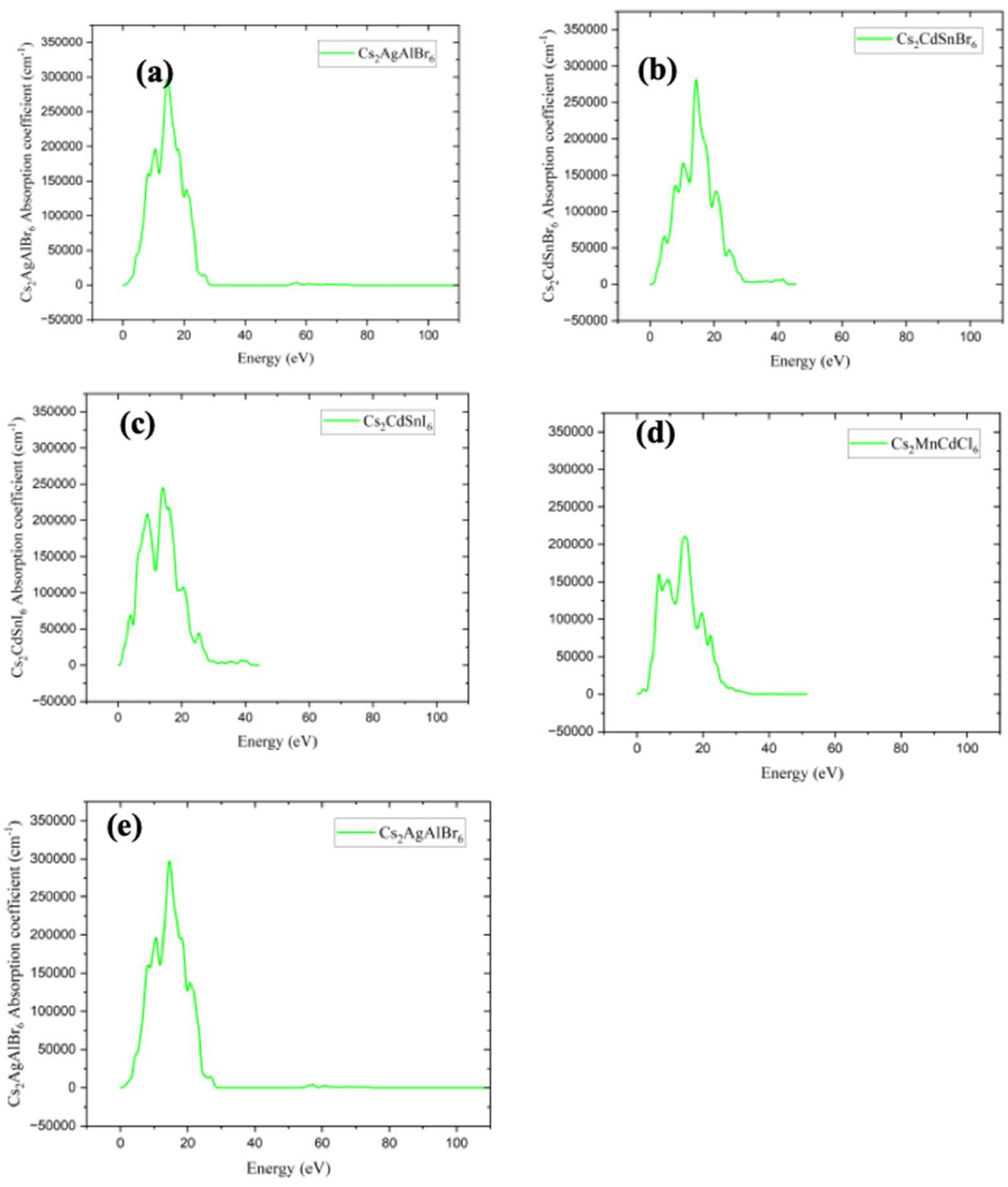

*Figure 15. Absorption spectra of the DFT-validated lead-free double perovskites. Calculated absorption coefficient as a function of photon energy is shown for the shortlisted compounds: (a) $Cs_2CdSnBr_6$, (b) $Cs_2CdSnI_6$, (c) $Rb_2SnMnBr_6$, (d) $Cs_2KGaI_6$, and (e)*

*$Cs_2AgAlBr_6$. All materials show strong inter-band absorption, with peak values on the order of $10^5$ $cm^{-1}$. This suggests good light-harvesting performance at the screening level. The differences in absorption onset, peak height, and spectral shape are mainly due to changes in band-edge chemistry, metal–halide hybridization, and dielectric response across the compounds.*

### 3.7.3 Refractive index and reflectivity

The refractive index n(ω) and reflectivity R(ω) translate dielectric response into optical impedance, which is important for device integration (see Table 12). Higher n(0) enhances light confinement but increases Fresnel reflection, making these properties critical for optical-stack design.

*Table 12. Refractive-index and reflectivity metrics.*

| Material | n(0) | n_max | k_max | R(0) | R_max |
|---|---|---|---|---|---|
| $Rb_2SnMnBr_6$ | 2.48 | 2.60 | 1.0 | 0.17 | 0.24 |
| $Cs_2CdSnBr_6$ | 2.27 | 2.50 | 1.2 | 0.16 | 0.31 |
| $Cs_2CdSnI_6$ | 2.86 | 3.00 | 1.5 | 0.36 | 0.37 |
| $Cs_2KGaI_6$ | 2.14 | 2.40 | 1.5 | 0.13 | 0.29 |
| $Cs_2AgAlBr_6$ | 2.14 | 2.30 | 1.3 | 0.14 | 0.34 |

Consistent with the dielectric trends, $Cs_2CdSnI_6$ shows the largest low-energy refractive index, whereas $Cs_2AgAlBr_6$ and $Cs_2KGaI_6$ occupy the lower-screening side of the validated set, as depicted in Figure 16. Reflectivity remains moderate for all candidates, which is favorable for absorber applications because a large fraction of the incident flux can still enter the film. The spectral structure of R(ω) also differentiates compositions that may be better suited to sensor optics or interface-engineered stacks rather than absorber-dominant roles.

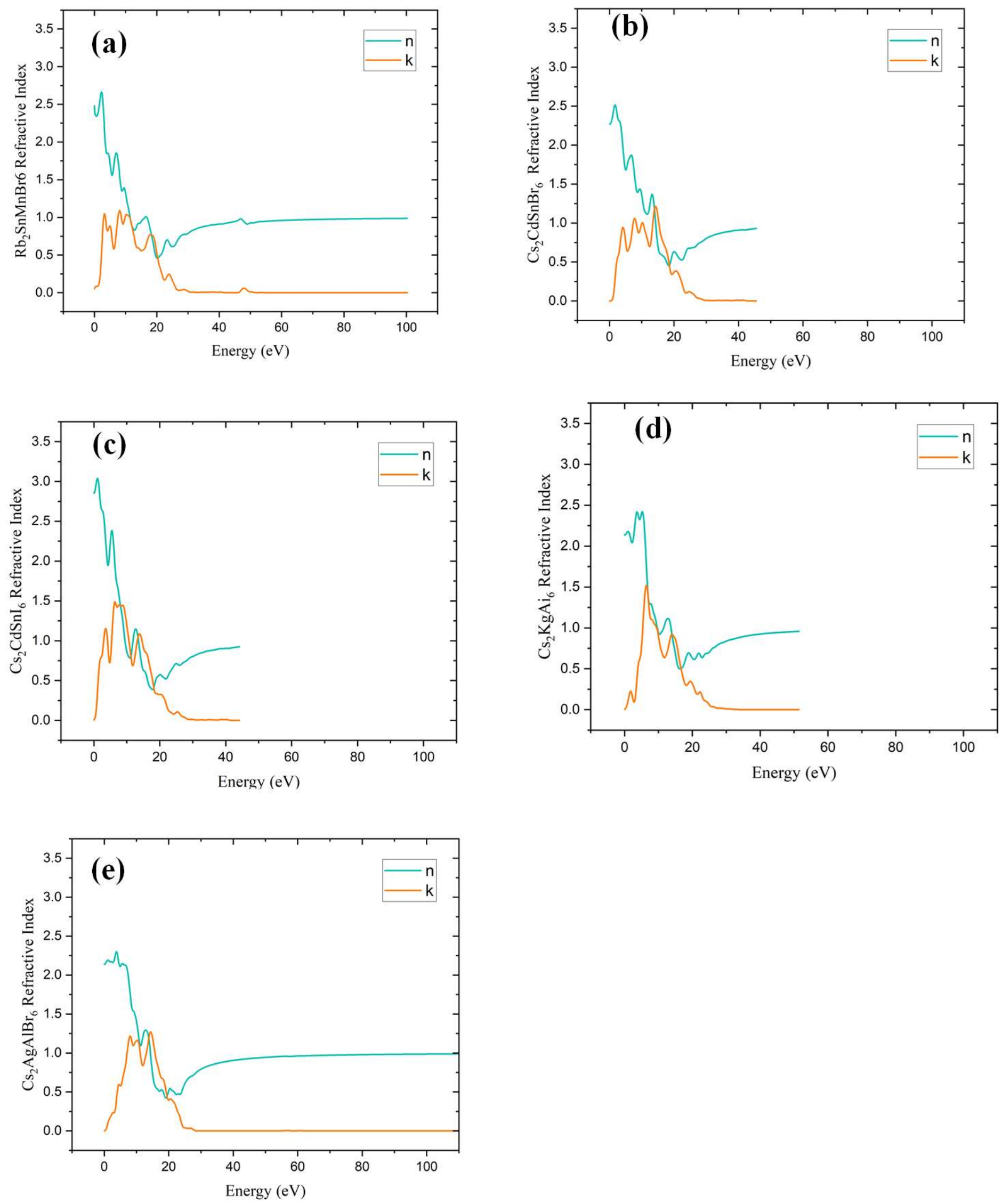

*Figure 16. Refractive index and extinction coefficient spectra of the studied double perovskites. This figure shows the real refractive index (n) and extinction coefficient (k) as a function of photon energy for the selected lead-free double perovskites: (a) $Rb_2SnMnBr_6$, (b) $Cs_2CdSnBr_6$, (c) $Cs_2CdSnI_6$, (d) $Cs_2KGaI_6$, and (e) $Cs_2AgAlBr_6$. All compounds show a strong optical response at low photon energies. However, the values of n and k vary from one material to another, showing different optical behaviors across the set.*

These properties describe how light travels through the materials and how much light is lost through reflection and absorption (Figures 16 and 17). Here, n(0) is the refractive index at zero photon energy. It shows the low-energy optical response. n_max is the highest refractive index value. k_max is the maximum value of the extinction coefficient. R(0) is the reflectivity at zero photon energy. R_max is the maximum reflectivity.

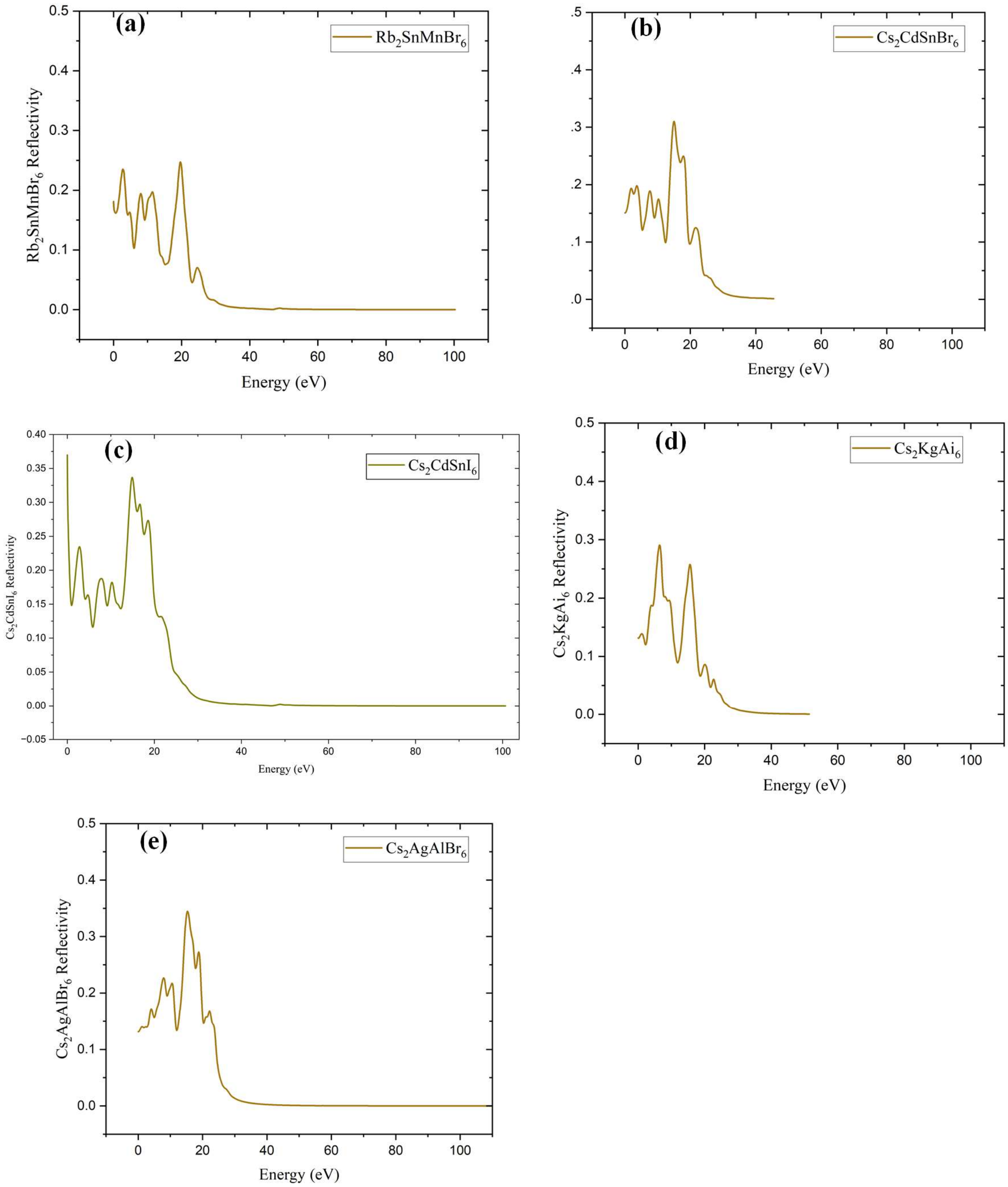

*Figure 17. Reflectivity spectra of the DFT-validated lead-free double perovskites. Calculated reflectivity as a function of photon energy is shown for $Rb_2SnMnBr_6$, $Cs_2CdSnBr_6$, $Cs_2CdSnI_6$, $Cs_2KGaI_6$, and $Cs_2AgAlBr_6$. All materials show moderate reflectivity at low photon energies. The peak positions and shapes change from one compound to another. These changes come from differences in dielectric constant, optical impedance, and band-edge chemistry across the materials.*

### 3.7.4 Loss function: excitation-energy fingerprints

The energy-loss function, L(ω) = Im[−1/ε(ω)] provides insight into collective electronic excitations and plasmon behavior. All compounds exhibit distinct loss peaks in the low- to mid-energy range, corresponding to bulk plasmon resonances (Figure 17). From the genome guided perspective, these optical features are mainly controlled by valence electron concentration, polarizability, and bonding nature. $Cs_2CdSnBr_6$ and $Cs_2CdSnI_6$ show sharper and stronger plasmon peaks around 18–22 eV. This suggests stronger collective electron excitation. It is linked to higher polarizability and better orbital overlap in these materials. In contrast, $Rb_2SnMnBr_6$ and $Cs_2KGaI_6$ show broader plasmon peaks. This indicates weaker coherence in the electronic excitation process. $Cs_2AgAlBr_6$ exhibits a shifted dominant peak near ~23 eV, suggesting a distinct balance of electron density and bonding. Overall, the loss spectra provide excitation fingerprints that reinforce the genome-property-performance relationship and extend validation to dynamic electronic response.

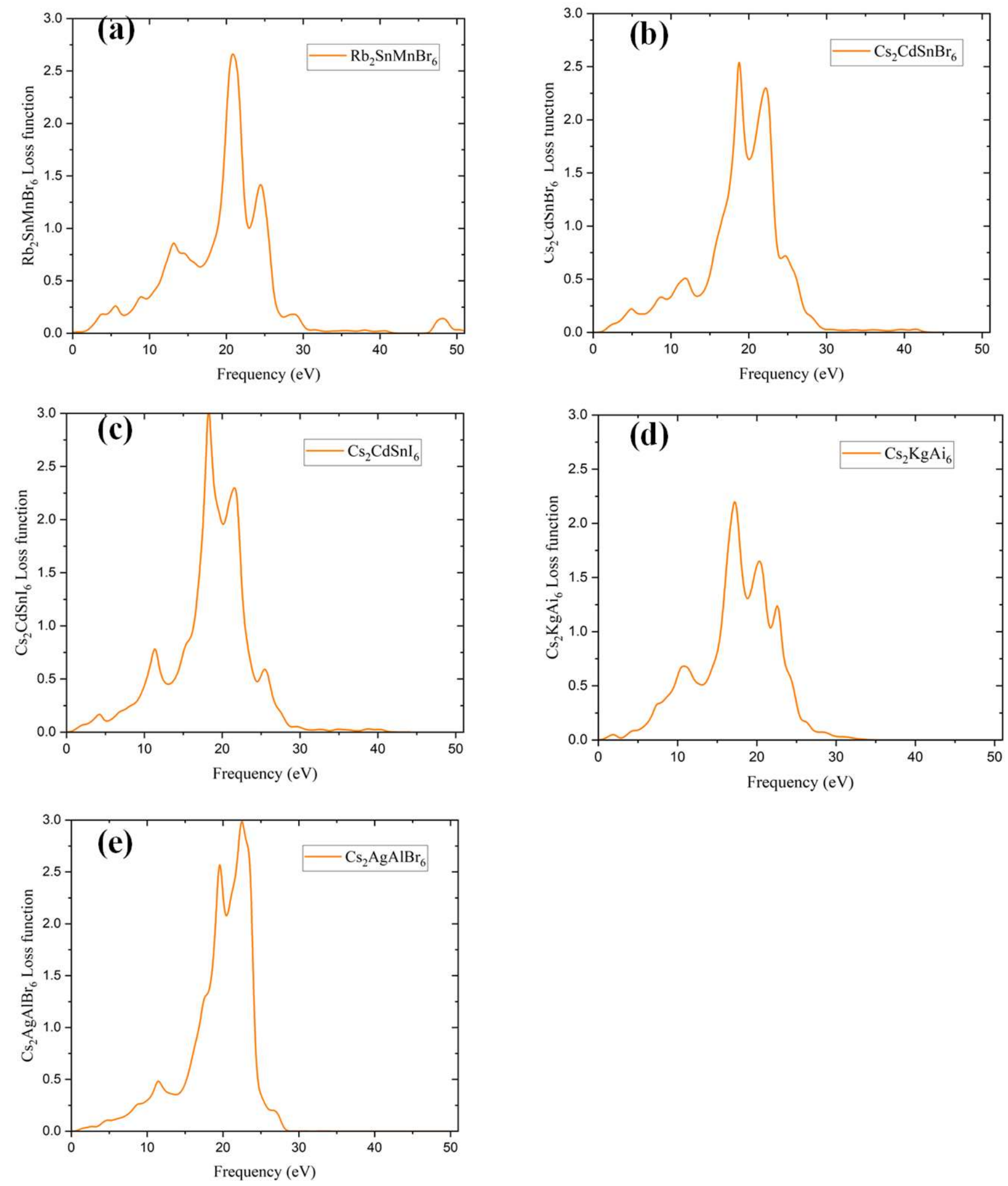

*Figure 17. Frequency-dependent energy-loss spectra of the shortlisted lead-free double perovskites. The dominant plasmon peaks provide excitation-energy fingerprints that complement the dielectric and absorption analyses.*

Plasmon resonance energies obtained from the main peaks of the calculated energy-loss function for the studied lead-free double perovskites recorded in Table 13. These values represent the main collective excitation energies obtained from the loss-function spectra.

*Table 13. Plasmon energies from the loss-function spectra.*

| Material | Minor plasmon peak (eV) | Intermediate peak (eV) | Dominant plasmon peak (eV) | Secondary strong peak (eV) |
|---|---|---|---|---|
| $Rb_2SnMnBr_6$ | 13.5 | 18.5 | 20.8–21.0 | 24.5 |
| $Cs_2CdSnBr_6$ | 14.0 | 17.8 | 18.8–19.0 | 21.8–22.0 |
| $Cs_2CdSnI_6$ | 12.0 | 17.5 | 18.5–18.8 | 21.0–21.5 |
| $Cs_2KGaI_6$ | 11.5 | 17.0 | 18.0–18.2 | 21.0–21.3 |
| $Cs_2AgAlBr_6$ | 12.0 | 19.8 | 22.5–22.8 | 23.5–24.0 |

### 3.7.5 Optical conductivity: transition-strength phenotype

Optical conductivity $\sigma(\omega)$ shows how strongly a material responds to light and how easily electrons are excited. It reflects the strength of photon-driven electronic transitions. All selected compounds show clear conductivity features at low to mid photon energies (below 25 eV), which indicates strong inter-band transitions (see Figure 18). $Cs_2CdSnBr_6$ and $Cs_2CdSnI_6$ show the strongest and widest conductivity response. This means they have a higher probability of electronic transitions and more efficient carrier generation. This behavior is related to good metal–halide hybridization, higher polarizability, and balanced bonding. In contrast, $Rb_2SnMnBr_6$ shows weaker conductivity because the valence states are more localized. $Cs_2KGaI_6$ and $Cs_2AgAlBr_6$ show intermediate behavior, with less balanced and less uniform transition profiles.

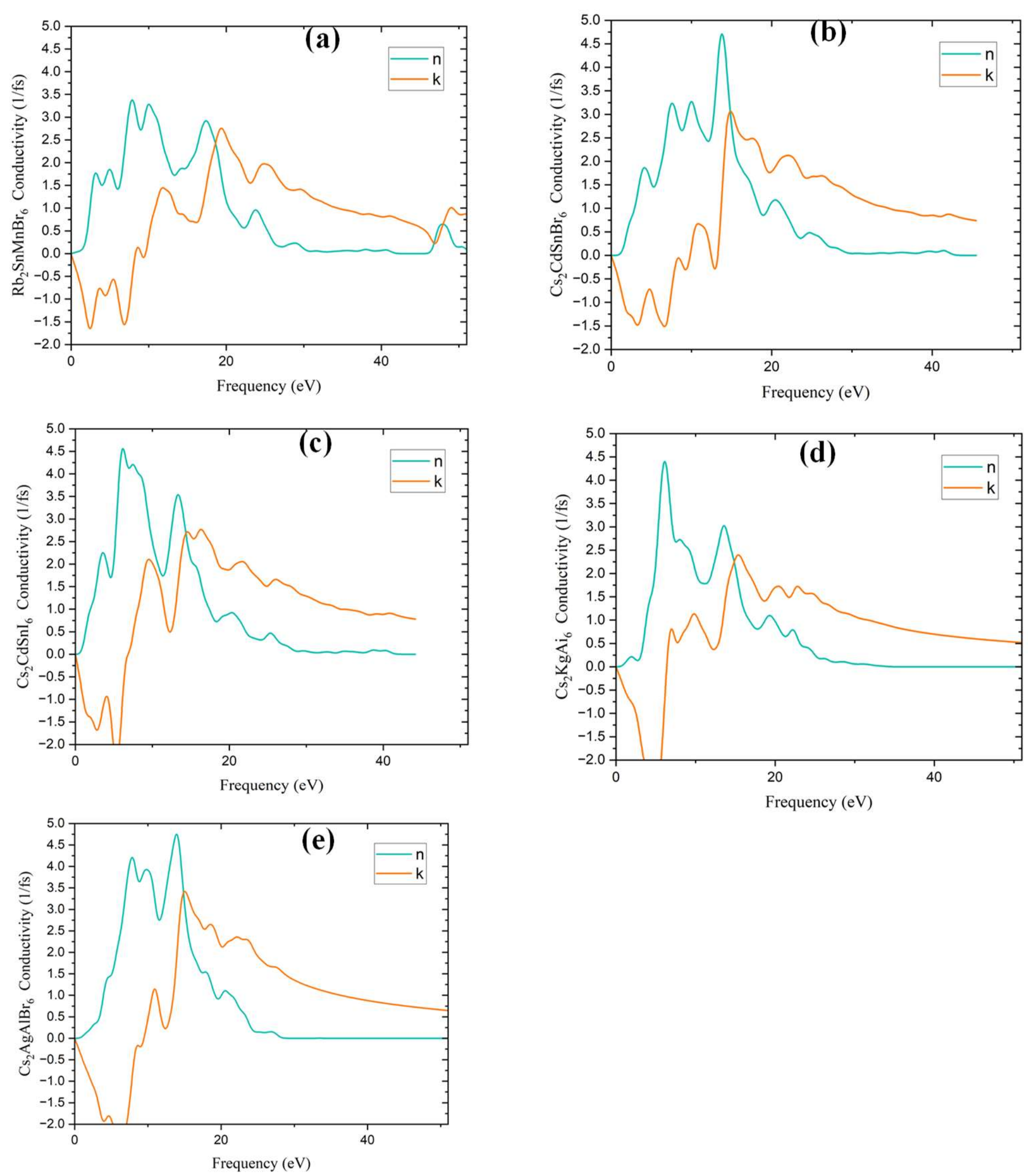

*Figure 18. Optical conductivity spectra of the studied double perovskites. This figure shows the optical conductivity as a function of photon energy for (a) $Rb_2SnMnBr_6$, (b) $Cs_2CdSnBr_6$, (c) $Cs_2CdSnI_6$, (d) $Cs_2KGaI_6$, and (e) $Cs_2AgAlBr_6$. All compounds show clear conductivity features in the low to mid energy range. This indicates strong inter-band electronic transitions. The differences in peak height and shape come from changes in orbital hybridization, dielectric response, and transition strength across the materials.*

The conductivity trends can be explained by basic material features. When there is strong orbital overlap, good mixing of band edges, and a higher dielectric response, the optical conductivity $\sigma(\omega)$ increases. When the electronic states are more localized and the bands are flatter, the conductivity becomes weaker. The Cd/Sn-based compounds show the highest $\sigma_1$,max values and also maintain a stronger response over a wider energy range (see Table 14). This supports their suitability as good absorber materials. The imaginary component of conductivity shows corresponding dispersive behavior, confirming dynamic carrier response. At higher photon energies (>30–40 eV), conductivity decreases, indicating reduced transition probability.

*Table 14. Peak optical-conductivity values derived from the calculated spectra.*

| Material | $\sigma_1$,max (1/fs) | E_peak (eV) | Secondary peak (eV) | High-energy tail behavior |
|---|---|---|---|---|
| $Rb_2SnMnBr_6$ | 3.3 | 9–10 | 18–20 | Gradual decay above 25 eV |
| $Cs_2CdSnBr_6$ | 4.7 | 12–13 | 18–22 | Smooth decay up to 40 eV |
| $Cs_2CdSnI_6$ | 4.5 | 8–9 | 15–18 | Moderate decay, broad tail |
| $Cs_2KGaI_6$ | 4.3 | 7–8 | 16–20 | Faster decay after 25 eV |
| $Cs_2AgAlBr_6$ | 4.6 | 10–12 | 16–19 | Broad tail up to ~40 eV |

Overall, the optical conductivity results describe how strongly electrons respond to light. This complements the dielectric and absorption behavior. It shows how easily electronic transitions occur in each material. The results also show that changes in material descriptors control both static optical properties and dynamic excitation processes. This supports the idea that the genome-guided screening approach is effective and physically meaningful.

### 3.8 DFT-Validated Design Rules from Descriptor–Phenotype Coupling

Inverse design is often reduced to model-driven screening, which can identify viable candidates but rarely provides actionable guidance on how to tune composition toward simultaneous stability and optical performance. Here, we extend inverse design beyond filtering by explicitly linking genome gene clusters (see Table 3) to DFT-derived optical phenotypes ($\varepsilon(\omega)$, $\alpha(\omega)$, $n(\omega)$, $R(\omega)$, $L(\omega)$, and $\sigma(\omega)$), thereby constructing a mechanistic genotype–phenotype map. In this framework, descriptors act as genes and DFT observables as phenotypes, enabling direct translation of model interpretability into generalizable design rules across the $A_2BB'X_6$ chemical space.

Compared with traditional high-throughput methods that rely on black-box models or single-property screening, this approach is more structured. It evaluates materials in separate but connected steps. The framework evaluates materials based on structural feasibility, thermodynamic stability, and optoelectronic behavior. Each layer is described using clear physical descriptors.

#### 3.8.1 Genome structure and modular control of phenotypes

The stability genome is divided into three clear and interpretable groups of descriptors (Table 3). The first group is packing genes, which include ionic radii and formability factors such as $t$, $\mu$, and $\tau$. The second group is bonding genes, which describe bond strength and electronegativity differences. The third group is Opto-electronic response genes, which include polarizability, ionization energy, electron affinity, valence electron count and atomic number. Correlation analysis (see Figure 3) shows that different descriptor groups control different properties. Stability and band gap are influenced by separate sets of descriptors. Band-gap variation is dominated by bonding and B-site response descriptors (e.g., B–X bond energy), whereas thermodynamic stability is influenced more strongly by broader thermochemical and A-site-related variables. This modularity motivates a sequential inverse-design strategy, where gene clusters act as ordered control layers: formability-stability-optical transition strength. Figure 19, along with Tables 16 and 17, makes this idea practical. It connects Genotype–phenotype coupling maps to DFT-derived optical properties.

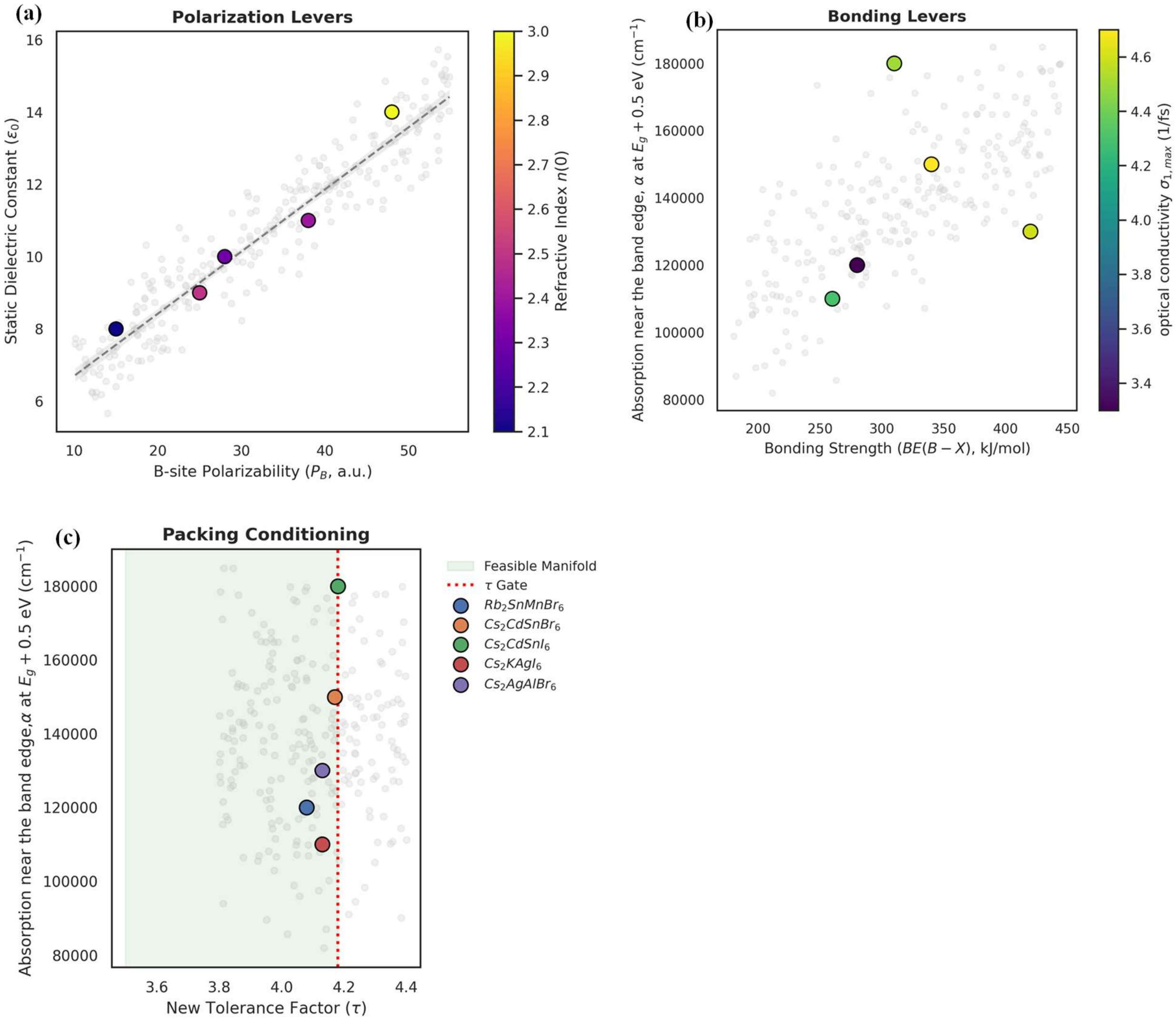

*Figure 18. Genotype–phenotype coupling maps linking interpretable machine-learning gene clusters to DFT-derived optical properties. (a) Polarization gene score versus dielectric constant ($\varepsilon_0$) and refractive index n(0). (b) Bonding gene score versus near-edge transition strength, represented by α(E_g + 0.5 eV) and $\sigma_1$(E_g + 0.5 eV). (c) Packing descriptor τ versus absorption-related phenotypes, highlighting structural conditioning within the formable manifold. Above-onset metrics are evaluated at E_g + 0.5 eV.*

### 3.8.2 Opto-electronic response genes: dielectric constant phenotypes and orbital-level band-edge interpretation

Opto-electronic response genes encode lattice and electronic susceptibility through atomic polarizability, ionization energy, and electron affinity. Their primary DFT manifestations are the static dielectric constant $\varepsilon_0$ and low-energy refractive index n(0), which control exciton screening and device readiness. They also determine the orbital character of valence- and conduction-band edges and therefore control how bonding advantages translate into optical transitions.

As shown in Figure 19(a), higher polarization-gene scores correlate directly with stronger dielectric screening. Within the shortlisted set, $Cs_2CdSnI_6$ exhibits the highest $\varepsilon_0$ and n(0), while $Cs_2KGaI_6$ and $Cs_2AgAlBr_6$ define the lower-screening limit, consistent with trends observed in Section 3.7. Their DFT manifestation is observed in the band structures and PDOS discussed in Section 3.6.3, which distinguish halide-p, metal-s, and metal-d contributions to band edges.

### 3.8.3 Bonding genes and near-edge transition strength

Bonding genes control both thermodynamic stability and optical transition strength through metal–halide bond energy and electronegativity contrast. These descriptors influence inter-band absorption $\alpha(\omega)$, optical conductivity $\sigma_1(\omega)$, and dielectric response $\varepsilon_2(\omega)$. Mechanistically, stronger B–X and B′–X bonding stabilizes the octahedral framework while simultaneously enhancing metal–halide hybridization at the band edges. This dual role explains why stability selection can co-select strong optical absorption when bonding descriptors are optimized within the formable region.

To enable direct comparison across materials, Figure 19(b) evaluates near-edge metrics at Eg + 0.5 eV. In this representation, $Cs_2CdSnBr_6$ and $Cs_2CdSnI_6$ show the strongest coupling between bonding strength and transition intensity, whereas $Rb_2SnMnBr_6$, $Cs_2KGaI_6$, and $Cs_2AgAlBr_6$ exhibit weaker above-onset response.

### 3.8.4 Packing genes as structural conditioning factors for optical onset

Packing genes are typically used as binary filters, but within the formable region they exert continuous control over octahedral geometry, orbital overlap, and band dispersion. Figure 19(c) shows that $\tau$ acts as a structural conditioning parameter for near-edge optical response. The shortlisted compounds occupy a narrow $\tau$ window (~4.08–4.18), defining a shared formability manifold. Within this region, $\tau$ does not directly determine absorption magnitude but modulates how effectively bonding and polarization descriptors translate into near-edge $\alpha$ and $\sigma_1$. This explains why some compositions with strong high-energy absorption exhibit weaker near-edge response.

3.8.5 Design-rule payoff of the genome-guided screening framework

The payoff of genome decoding is therefore not simply faster ranking. It is the ability to convert descriptor interpretability into sequential design rules: first satisfy packing feasibility, then stabilize the framework through bonding chemistry, and finally tune dielectric constant and transition strength through polarization and band-edge identity. In this sense, the workflow moves beyond black-box screening and becomes a mechanistic inverse-design engine that can be transferred to wider $A_2BB'X_6$ chemistries.

Table 15 showed the summary of how the stability-genome descriptor groups are expressed in the DFT-derived structural, optical, and electronic phenotypes of the shortlisted $A_2BB'X_6$ absorbers. Packing descriptors define the structural manifold, bonding descriptors govern framework stabilization and near-edge transition strength, and optoelectronic-response descriptors control dielectric constant, band-edge character, and dispersion-related optical behavior. Whereas Table 16 summarized the design-rule summary translating gene clusters into mechanistic understanding and actionable inverse-design controls.

*Table 15. Descriptor–phenotype coupling with quantitative DFT evidence. Packing descriptors define the structural manifold; bonding descriptors govern stability and near-edge transition strength; optoelectronic-response descriptors control dielectric screening and band-edge behavior.*

| Gene cluster | Gene descriptors (Table 3) | Primary DFT phenotype(s) used in this work | Phenotype evidence from the shortlisted perovskites |
|---|---|---|---|
| Packing/ formability | Ionic radii, tolerance factors t, μ, and τ | Structural feasibility; conditioning of near-edge response; dispersion-sensitive characteristics | The shortlisted absorbers lie within a narrow formable window, with τvalues of about 4.08-4.18, indicating a shared structural manifold for optical optimization. This shows that packing descriptors define the first structural condition for realizing stable and optically useful compositions. |
| Bonding/ framework cohesion | Bond dissociation energies, electronegativity differences; atomic formation enthalpies | Absorption $\alpha(\omega)$, optical conductivity $\sigma_1(\omega)$, inter-band dielectric response $\varepsilon_2(\omega)$, and $E_{hull}$-based stability condition | At $E_g + 0.5$eV, the Cd/Sn compounds show the strongest near-edge coupling, whereas $Rb_2SnMnBr_6$, $Cs_2KGaI_6$, and $Cs_2AgAlBr_6$ show weaker above-onset response. This suggests that stronger metal–halide bonding and better orbital hybridization help both thermodynamic stability and stronger optical transitions near the absorption edge. |

| Opto-electronic response | Ionization energy; electron affinity; polarizability; valence electron count, atomic number | Static dielectric constant $\varepsilon_0$, refractive index n(0), dispersive n(ω), band structures, PDOS, onset character, and band-dispersion indicators | $Cs_2CdSnI_6$ shows the highest screening response, with $\varepsilon_0 \approx 8.16$and $n(0) \approx 2.86$, while $Cs_2KGaI_6$ and $Cs_2AgAlBr_6$ define the lower-screening edge of the validated set. Band structure and PDOS further explain why strong high-energy absorption does not always produce strong near-edge $\sigma_1$, as seen for $Cs_2AgAlBr_6$ and $Cs_2KGaI_6$ relative to the Cd/Sn absorbers. |
|---|---|---|---|

*Table 16. DFT-validated design rules from the genome guided framework. Gene clusters are translated into mechanistic interpretations, actionable tuning strategies, and expected device-level impact.*

| Gene cluster | Mechanistic interpretation (genotype to phenotype) | Actionable inverse-design lever | Expected device impact |
|---|---|---|---|
| Packing/ formability | Ionic-size matching and octahedral geometry define whether the $A_2BB'X_6$ framework can form. Within the formable region, t, μ, and τ also influence orbital overlap and near-edge optical response. | Enforce t/μ/τ feasibility first, then optimize A-, B-, and B′-site size matching to minimize distortion. | Higher structural realizability, lower distortion/trap tendency, and better optical/transport conditioning. |
| Bonding/ framework cohesion | Metal-halide bond strength, electronegativity contrast, and chemical balance govern framework stability and band-edge hybridization. Stronger B–X/B′–X bonding supports both stability and near-edge transition strength. | After structural screening, favor stronger B–X/B′–X bonding proxies and balanced electronegativity differences within the target E_g window. | Greater thermodynamic stability leads to easier material formation, stronger absorption near the band edge, and better performance in thin-film absorber applications. |

| Opto-electronic response | The optoelectronic response is mainly controlled by polarizability, ionization energy, electron affinity, valence electron count, and atomic number. These factors determine the dielectric constant, band-edge nature, and carrier imbalance in the material. | After stability screening, the next step is to tune the halide type and B/B′ site chemistry. This helps improve the dielectric constant and adjust the band-edge characteristics. These changes are verified using band structure and PDOS analysis. | Help to reduce excitonic effects, improve charge separation, and provide better control over the band gap. As a result, the materials become more suitable for photovoltaics, sensors, and other optoelectronic applications. |
|---|---|---|---|

### 3.8.7 Practical design rules for phase-stable and optically functional $A_2BB'X_6$ absorbers

The descriptor–phenotype analysis gives a compact set of practical design rules for lead-free $A_2BB'X_6$ absorbers. These rules should not be taken as strict universal laws. Rather, they are physically guided screening principles derived from the combined machine-learning and DFT results of the present genome-guided framework.

First, structural feasibility must be satisfied before functional optimization. The packing descriptors, especially ionic radii and the formability metrics t, μ, and τ, define the structural manifold where the double-perovskite framework can exist. If a composition is outside this region, later optimization of electronic or optical properties becomes much less meaningful. Second, framework cohesion should be optimized inside the formable region. Stronger B–X and B′–X bonding, together with balanced electronegativity differences, are associated with higher thermodynamic accessibility and stronger near-edge optical response. This means that metal-halide bonding acts not only as a stability factor, but also as a functional design handle. Third, dielectric constant and band-edge behavior should be tuned through the optoelectronic-response descriptors. Polarizability, ionization energy, electron affinity, valence electron count, and atomic number influence how the realized structure expresses screening, charge redistribution, and orbital-level response. In the validated set, this is seen clearly in the stronger screening of $Cs_2CdSnI_6$ compared with lower-screening compounds such as $Cs_2KGaI_6$ and $Cs_2AgAlBr_6$.

Finally, band structure and PDOS should be used as the orbital-level check of the screening logic. The descriptor genome identifies promising regions of composition space, but the physical interpretation becomes much stronger when the DFT electronic structure confirms the expected band-edge chemistry.

### 3.9 Application mapping as a genome guided-design rule output: a portfolio perspective

A practical inverse-design workflow should return a device-matched portfolio rather than a single universal best material. The validated materials fall into different regions of band gap, screening strength, and transition intensity. Because of this, they show different optical and electronic behaviors. This means each compound can be suited for different optoelectronic applications.

$Cs_2CdSnBr_6$ and $Cs_2CdSnI_6$ define the absorber-grade frontier of the validated set because they combine application-relevant gaps with the strongest above-onset absorption and the most balanced optical phenotype suite. $Cs_2CdSnI_6$ further benefits from the largest dielectric constant, which is favorable for exciton dissociation. $Cs_2KGaI_6$ occupies the narrow-gap edge of the portfolio and is better aligned with near-infrared sensing, while $Rb_2SnMnBr_6$ and $Cs_2AgAlBr_6$ are more naturally interpreted as asymmetric-transport systems whose band-edge chemistry may be useful in auxiliary optoelectronic roles rather than as the primary photovoltaic absorber. Figure 20 and Table 17 summarize this mapping by combining band gap, near-edge absorption, dielectric constant, and transport characteristics into an application-oriented comparison.

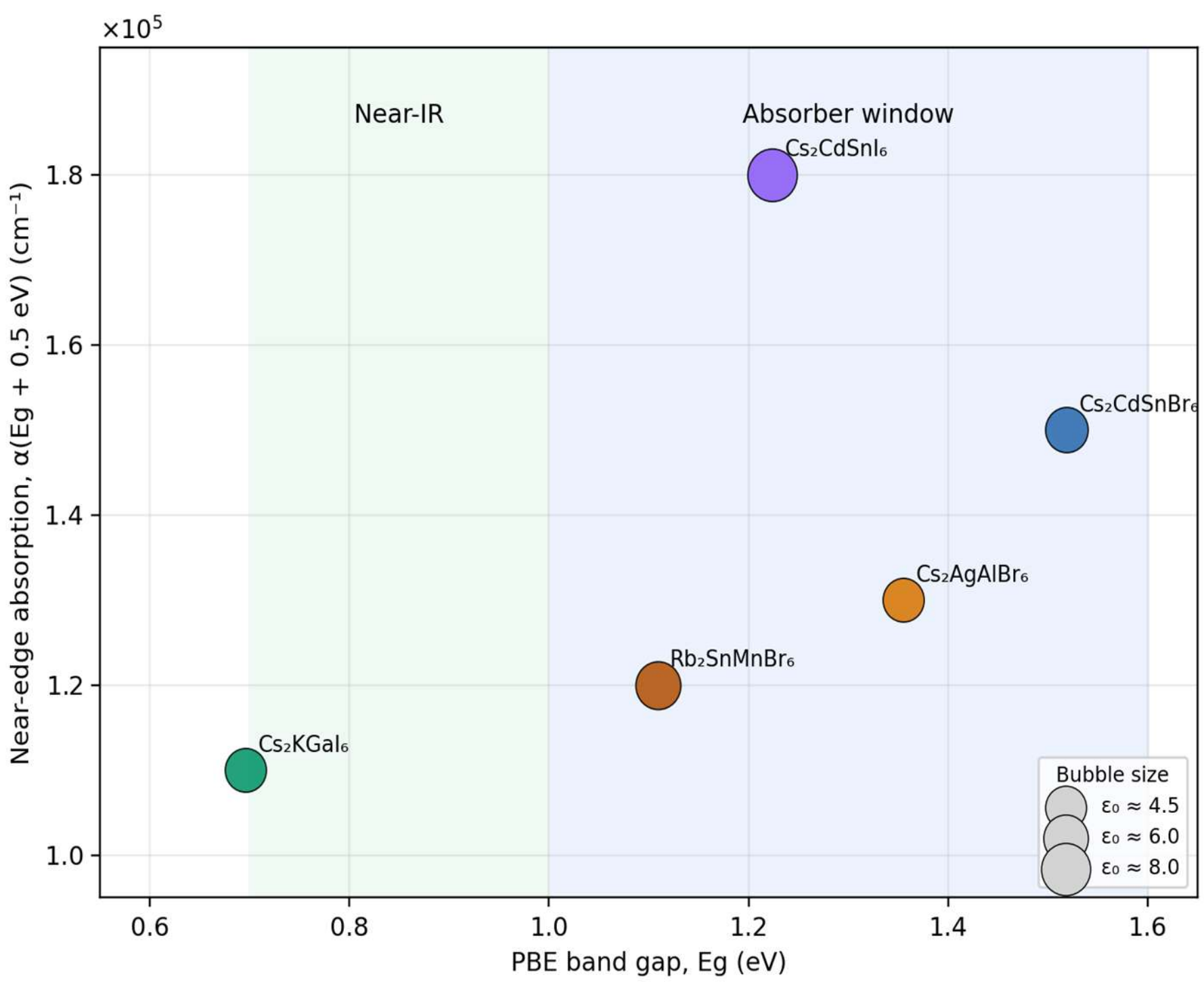

*Figure 19. Application-oriented map of the DFT-validated candidates. The x-axis shows the PBE band gap, the y-axis shows α(Eg + 0.5 eV), and bubble size scales with $\varepsilon_0$. The Cd/Sn systems define the absorber window, whereas $Cs_2KGaI_6$ occupies the near-IR edge.*

*Table 17. Application mapping of the DFT-validated candidates with caveats. This table is presented as an application map, not as a final device ranking.*

| Compound | Screening-level role | Supporting strengths | Main limitation | Confidence level | Key caveat |
|---|---|---|---|---|---|

| $Cs_2CdSnBr_6$ | Promising absorber candidate | Largest PBE gap among top candidates in the desirable visible range; strong near-edge absorption; light carrier masses | Lower dielectric constant than $Cs_2CdSnI_6$ | Medium | Application assignment is based on PBE-level screening metrics and should be confirmed using HSE06 + SOC |
|---|---|---|---|---|---|
| $Cs_2CdSnI_6$ | Promising absorber / photodetector candidate | Strong dielectric constant; strong absorption; iodide-driven, red-shifted response | Heavier carriers; likely stronger SOC sensitivity | Medium | Absolute gap and rank may shift at higher level of theory |
| $Cs_2KGaI_6$ | Near-IR or narrow-gap optoelectronic candidate | Lowest PBE gap among shortlisted compounds; iodide-rich optical response | Flatter bands and weaker transport characteristics | Low to medium | Quantitative role assignment is especially sensitive to band-gap correction |
| $Rb_2SnMnBr_6$ | Asymmetric-transport or spin-active optoelectronic candidate | Electron-favored transport; Mn-driven distinctive band-edge character | Very heavy holes; magnetic-state dependence | Low to medium | Functional assignment depends on the preferred magnetic ground state |

| $Cs_2AgAlBr_6$ | Auxiliary optoelectronic / photodetector candidate | Strong high-energy absorption; relatively light electrons | More modest dielectric constant and weaker near-edge response | Medium | Better suited as a secondary rather than primary absorber candidate at the present screening level |
|---|---|---|---|---|---|

3.10 Scope and limitations of the present screening level

The present workflow is intentionally screening-oriented. All band gaps and optical spectra are referenced to the same scalar-relativistic PBE level without explicit spin-orbit coupling, hybrid-functional, or GW corrections, which means that absolute values for heavy-element systems should be interpreted cautiously. In addition, the present validation focuses on static phase stability, relaxed structures, and optoelectronic phenotypes; finite-temperature lattice stability, defect chemistry, explicit excitonic effects, and experimental synthesis remain outside the present scope. These omissions do not invalidate the relative ranking because the entire candidate set is treated on a consistent computational level, but they define the natural next validation layer before experimental translation.

4. Conclusions

We recast lead-free double-perovskite discovery as a stability-genome guided interpretable screening and design rule problem. The genome-guided ML-DFT workflow developed here decodes interpretable descriptor families, converts device constraints into compositions, and closes the loop with DFT verification of structure, stability, electronic structure, dielectric constant, absorption, excitation fingerprints, and optical conductivity. Across 13,088 enumerated $A_2BB'X_6$ compositions, this framework identifies five phase-stable semiconductor candidates with favorable band gaps, strong optical absorption, and meaningful dielectric constant. More importantly, it yields transferable design rules: packing genes define the formability manifold, bonding genes govern near-edge transition strength, and Opto-electronic response genes regulate dielectric constant and exciton readiness. In this sense, inverse design becomes more than accelerated screening; it becomes a mechanistic route for navigating chemical space and discovering stable, high-performance lead-free double perovskites for semiconductors, sensors, and photovoltaic applications.

# Supporting Information

## Genome-Guided Interpretable Screening of Phase-Stable, Lead-Free Double Perovskite Absorbers for All-Inorganic Semiconductors, Sensors, and Photovoltaics with DFT-Validated Design Rules

*Nafis Ahtasum*, Sohanur Rahman Sohan, Md. Mostaq Ahmed Himel, Md. Zahid Hassan, Muhammad Harussani Moklis, Masud Rana Rashel, Hasan Jamil, AKM Kamrul Islam, Mouhaydine Tlemcani*

Table S1 | Comparison with representative prior ML screening studies of lead-free double perovskites

| Study | Search space size | Properties predicted | Tuning method | Validation type | New candidates |
|---|---|---|---|---|---|
| **Im et al., (2019)** | 540 | Heat of formation and band gap (ML) | GBRT; feature-importance analysis | DFT-computed dataset with train/test evaluation | Design rules proposed; no explicit final shortlist reported |
| **Guo & Lin, (2021)** | 540 | Thermodynamic stability and band gap (ML) | Model comparison (RF, RR, SVR, XGBoost) | First-principles dataset with train/test evaluation | Potential candidates discussed; no explicit final shortlist reported |
| **Gao et al., (2021)** | 5,796 | Band gap (ML); stability and optical-property follow-up (DFT) | XGBR; model comparison | Combined ML screening and DFT validation | 2 novel lead-free inorganic double perovskites |
| **Liang & Zhang, (2022)** | 469 | Thermodynamic stability classification + E<sub>hull</sub> regression | Six tree-based ML algorithms; model selection | DFT-calculated E<sub>hull</sub> dataset | Screening model proposed; explicit final candidate count not visible in accessible abstracts |
| **Landini et al., (2022)** | 7,056 | Band gap (ML); follow-up stability, PCE, and effective masses (DFT) | Neural network | Hybrid-DFT evaluation of selected candidates | Novel potential absorbers proposed |
| **Wang et al., (2025)** | ~4,573 | Band gap and formation energy (ML) | Model comparison plus feature selection | Cross-validation and database cross-check | 99 predicted candidates, including 95 newly identified materials |
| This work | 13,088 | Stability (classification), band gap (regression), and geometry filters | Genetic-algorithm hyperparameter optimization | High-accuracy DFT on top-ranked shortlist | 24 shortlisted; 5 DFT-validated |

**Table S2. Optimized hyperparameters of the final machine-learning surrogate models**

| **Task** | **Model** | **Hyperparameter** | **Value** |
|---|---|---|---|
| **Stability classification** | Decision Tree | max_depth | 6 |
| **Stability classification** | Decision Tree | min_samples_split | 5 |
| **Stability classification** | Decision Tree | min_samples_leaf | 9 |
| **Stability classification** | Decision Tree | criterion | gini |
| **Stability classification** | Decision Tree | splitter | best |
| **Stability classification** | Decision Tree | max_features | 0.5699 |
| **Band-gap regression** | XGBoost | learning_rate | 0.0801 |
| **Band-gap regression** | XGBoost | max_depth | 8 |
| **Band-gap regression** | XGBoost | n_estimators | 380 |
| **Band-gap regression** | XGBoost | subsample | 0.9114 |
| **Band-gap regression** | XGBoost | reg_lambda | 0 |
| **Band-gap regression** | XGBoost | reg_alpha | 0.3901 |
| **Band-gap regression** | XGBoost | gamma | 0.3351 |
| **Band-gap regression** | XGBoost | colsample_bylevel | 0.9534 |
| **Band-gap regression** | XGBoost | colsample_bytree | 0.9277 |

**Table S2.** Optimized hyperparameters of the final surrogate models used for thermodynamic stability classification and PBE band-gap prediction.

**Table S3**. Class distribution (stable vs unstable) in the 1,221-compound dataset and 5-fold stratified cross-validation performance of the three stability classifiers. Metrics are reported as mean ± standard deviation over the five folds.

**(a) Class distribution in the 1,221-compound dataset**

| Class | Definition | Count | Fraction (%) |
|---|---|---|---|
| **Stable** | $E_{hull} \leq 25$ meV atom$^{-1}$ | 244 | 20.0 |
| **Unstable** | $E_{hull} \leq 25$ meV atom$^{-1}$ | 977 | 80.0 |
| **Total** | – | **1,221** | **100.0** |

**(b) Cross-validation performance (5-fold, stratified)**

Values reported as mean ± standard deviation.

| Model | Accuracy | Precision (stable) | Recall (stable) | ROC-AUC |
|---|---|---|---|---|
| **Random Forest** | 0.846 ± 0.015 | 0.835 ± 0.021 | 0.792 ± 0.025 | 0.933 ± 0.012 |
| **XGBoost** | 0.860 ± 0.012 | **0.865 ± 0.018** | 0.792 ± 0.022 | **0.959 ± 0.010** |
| **Decision Tree** | 0.833 ± 0.020 | 0.785 ± 0.025 | **0.831 ± 0.019** | 0.930 ± 0.022 |

**Table S4. Probability threshold selection and precision–recall trade-off**

*Effect of decision threshold on stable-class metrics for the EA-optimized Decision Tree. The bold row indicates the threshold adopted in the screening workflow to maximize recall.*

| Threshold on P(stable) | Precision (stable) | Recall (stable) | Notes |
|---|---|---|---|
| **0.50** | 0.785 | 0.831 | Default threshold |
| **0.45** | 0.762 | 0.855 | |
| **0.40** | 0.738 | 0.880 | |
| **0.35** | 0.705 | 0.915 | |
| **0.30** | **0.672** | **0.942** | **Selected (Max Recall)** |

**Table S5. Cross-validation performance of band gap regressors**

*Five-fold cross-validation performance on the 1,221-compound dataset (Mean ± Std).*

| Model | $R^2$ | RMSE (eV) | MSE ($eV^2$) |
|---|---|---|---|
| **SVR (RBF kernel)** | 0.9244± 0.035 | 0.5414± 0.052 | 0.2931± 0.045 |
| **Random Forest** | 0.9241± 0.028 | 0.5426± 0.041 | 0.2944± 0.038 |
| **XGBoost** | 0.9317± 0.015 | 0.5144± 0.022 | 0.2646± 0.018 |

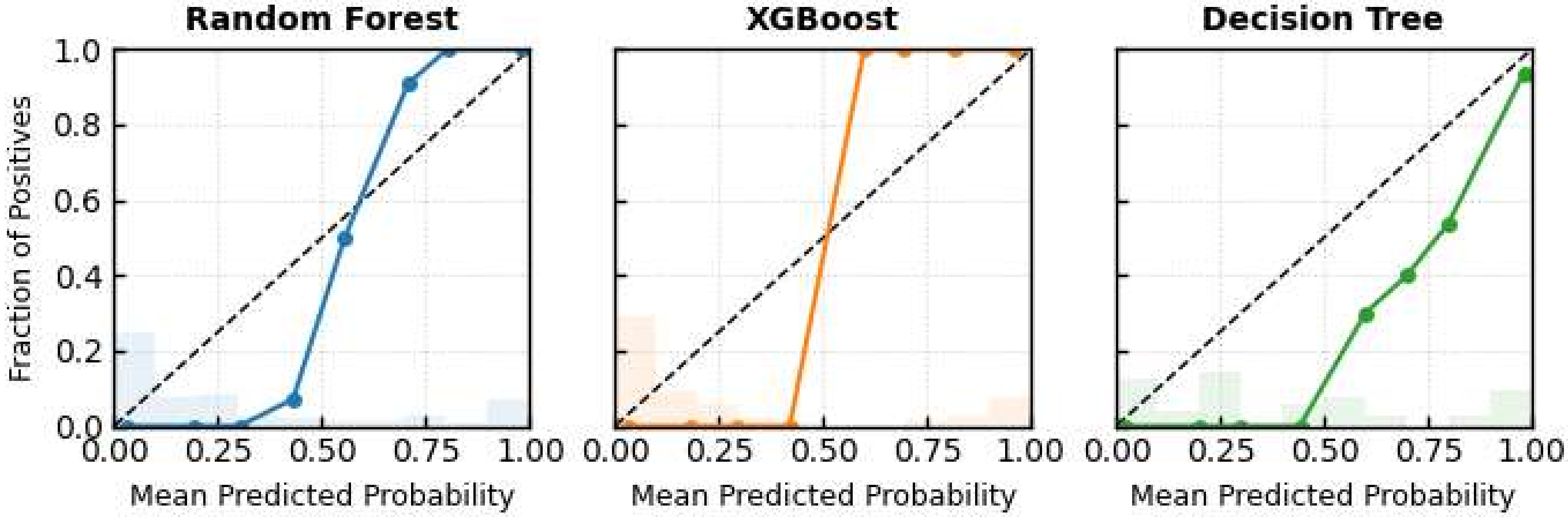

**Figure S1.** Reliability analysis of the three stability classifiers. (a) Calibration curve for the EA-optimized Random Forest classifier, comparing predicted P(stable) with empirical frequencies of the stable class in binned probability intervals. (b) Calibration curve for the EA-optimized XGBoost classifier. (c) Calibration curve for the EA-optimized Decision Tree classifier used in the screening pipeline. The dashed diagonal represents perfect calibration. The Decision Tree shows good calibration in the mid- to high-probability regime relevant for the screening threshold, supporting its use for probability-based selection of thermodynamically stable candidates.

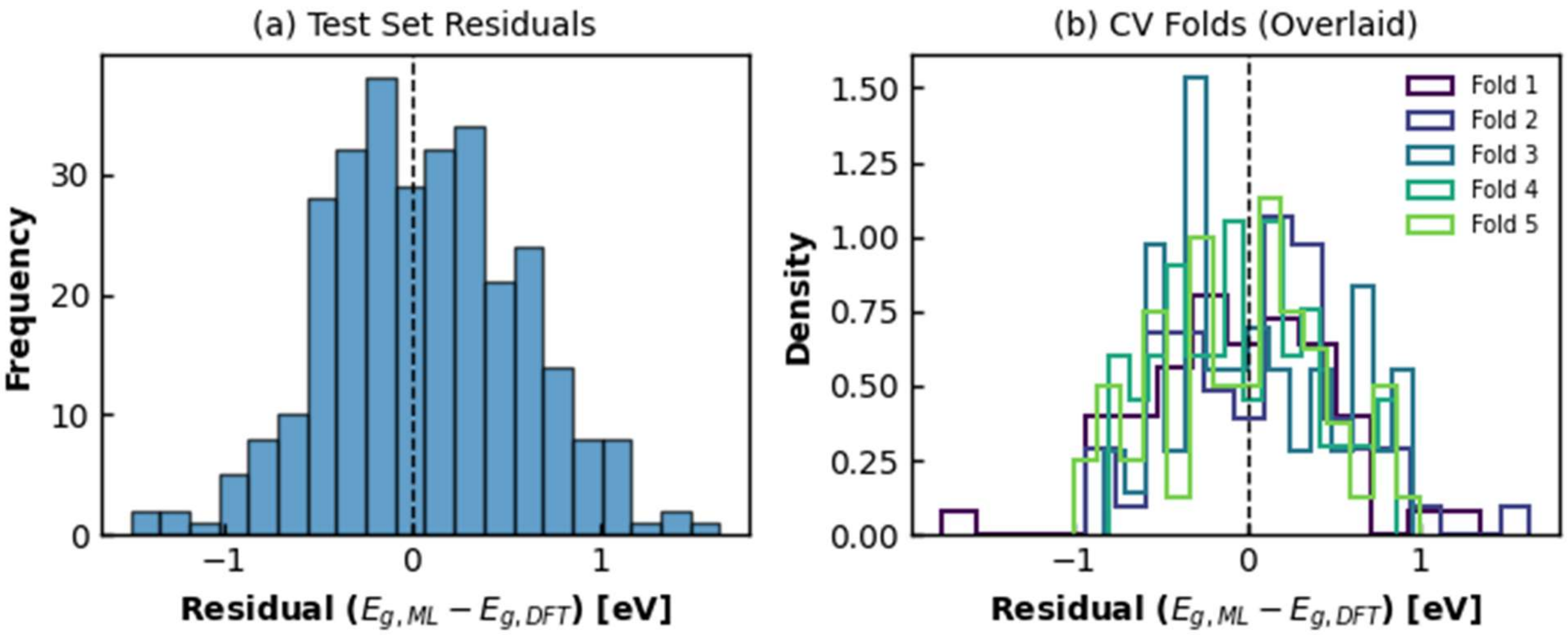

**Figure S2.** Distribution of residuals for ML-predicted band gaps (E_g,ML) relative to DFT values (E_g,DFT). (a) Histogram of residuals (E_g,ML – E_g,DFT) for the EA-optimized XGBoost regressor on the held-out test set. (b) Overlaid residual histograms for the five cross-validation folds, illustrating the stability and symmetry of the error distribution. The residuals are approximately centered at zero with no pronounced heavy tails, indicating the absence of strong

systematic bias or catastrophic outliers across the 0–8 eV band gap range.

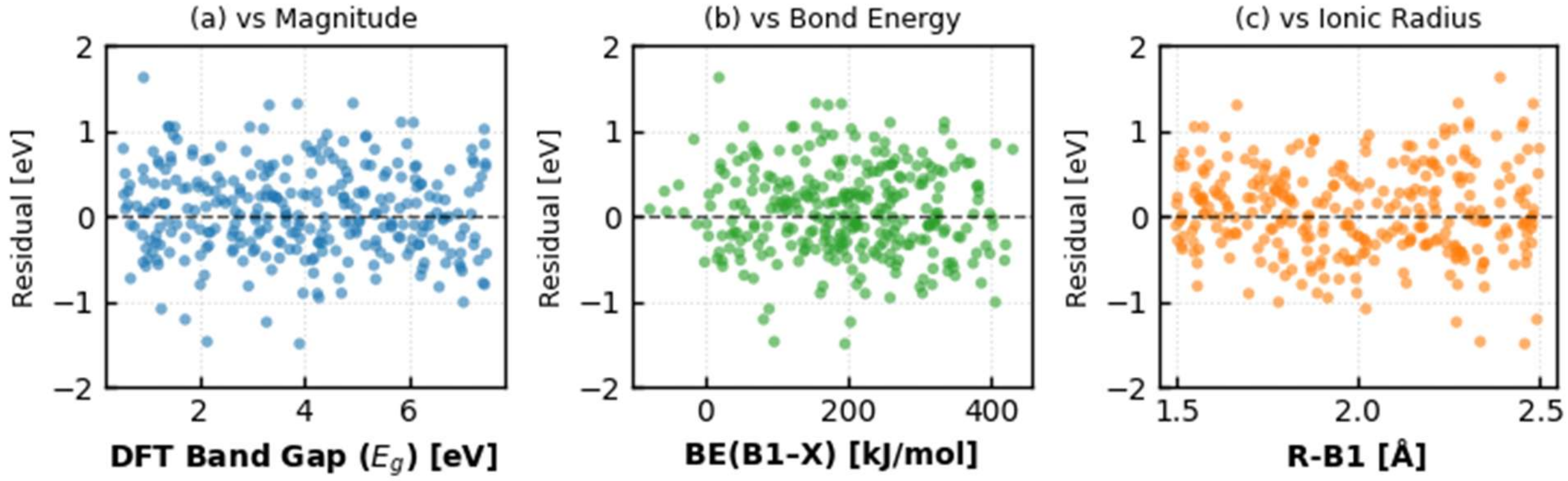

**Figure S3.** Dependence of band gap residuals on band gap magnitude and key descriptors for the EA-optimized XGBoost regressor. (a) Residuals (E_g,ML – E_g,DFT) as a function of DFT band gap E_g,DFT for all compounds in the test set. (b) Residuals as a function of the B–X bond dissociation energy BE(B1–X). (c) Residuals as a function of the B-site ionic radius R–B1. (d) Residuals as a function of B-site electronegativity X–B1. The absence of strong trends or clustering in these plots indicates that the model error does not systematically deteriorate for large band gaps or for particular regions of descriptor space, supporting the use of the XGBoost regressor as a reliable surrogate in the high-throughput screening workflow.

**Table S6. Convergence tests for DFT calculations**

To ensure numerical reliability, the convergence of the total energy ($E_{tot}$), lattice parameter (a), and band gap (Eg) was tested with respect to k-point sampling and basis-set quality. Tests were performed on representative nonmagnetic (Cs2SnMgBr6) and magnetic (Cs2MnSnBr6) systems. The final parameters (2×2×2 mesh, DNP basis) yielded results within negligible error margins relative to the computationally expensive 3×3×3 benchmark calculations.

Deviations are defined as:

$\Delta E = E - E^{(3\times3\times3)}$
$\Delta a = a - a^{(3\times3\times3)}$
$\Delta Eg = Eg - Eg^{(3\times3\times3)}$

where the superscript (3×3×3) denotes the benchmark calculation. A 2×2×2 k-point mesh was chosen as it offers an optimal balance between computational cost and accuracy.

(a) Example: $Cs_2SnMgBr_6$

| k-point mesh | Basis setting | ΔE (meV/atom) | Δa (Å) | ΔEg (eV) | Used in production |
|---|---|---|---|---|---|
| **1×1×1** | DNP (coarse) | 45.2 | 0.120 | 0.35 | No |
| **2×2×2** | DNP | 0.5 | 0.002 | 0.01 | Yes |
| **3×3×3** | DNP (fine) | 0 (benchmark) | 0 | 0 | No |

(b) Example: $Cs_2MnSnBr_6$

| k-point mesh | Basis setting | ΔE (meV/atom) | Δa (Å) | ΔEg (eV) | Used in production |
|---|---|---|---|---|---|
| **1×1×1** | DNP (coarse) | 52.8 | 0.145 | 0.41 | No |
| **2×2×2** | DNP | 0.9 | 0.004 | 0.02 | Yes |
| **3×3×3** | DNP (fine) | 0 (benchmark) | 0 | 0 | No |

Note: Deviations (Δ) are calculated relative to the 3×3×3 benchmark calculation. A 2×2×2 k-point mesh was chosen as it offers the optimal balance between computational cost and accuracy.

**Table S7. Elemental screening policy used in Phase III candidate filtering.**

| Category | Elements | Policy | Rationale |
|---|---|---|---|
| **Excluded: high-toxicity / regulatory concern** | Pb (always); As, Be, Cd (default) | Removed from the *practical* shortlist | Aligns with environmental safety and common regulatory constraints for scalable PV deployment. |
| **Controlled exception** | Cd (single case only) | Retained once as a diagnostic/benchmarking case; not claimed "non-toxic" | Allows comparison against chemically informative B-site combinations while keeping sustainability claims transparent. |
| **Excluded: noble / scarce** | Au, Pt, Ir (and other high-cost | Removed | High cost and limited supply make scaling |
| **(cost/supply risk)** | noble metals) | | unlikely for PV manufacturing. |
| **Excluded: elevated supply risk (criticality)** | [Insert your chosen list: e.g., In, Ga, Sb, etc.] | Removed when exceeding a defined criticality threshold | Reduces exposure to supply-chain volatility; improves scalability. |
| **Conditional / justified inclusion (not "earth-abundant")** | Te (example) | Allowed only when the target application benefits (e.g., IR absorber); labeled "less abundant but Pb-free" | Maintains functional breadth (IR band gaps) while avoiding over-claiming sustainability. |

**Table S8. Numeric thresholds used for geometric pre-filtering of $A_2BB'X_6$ compositions.**

| Metric | Definition (brief) | Acceptance window used in this work | Notes |
|---|---|---|---|
| **Goldschmidt tolerance factor $t$** | $t = \frac{r_A + r_X}{\sqrt{2}(r_B + r_X)}$ (using effective $r_B$ for B/B′) | **$t$: 0.80–1.10** | Radii from Shannon; $r_B$ taken as average of B and B′ radii for screening. |
| **Octahedral factor $\mu$** | $\mu = \frac{r_B}{r_X}$ (effective $r_B$) | $\mu$: **0.41–0.90** | Ensures $BX_6$ octahedral formability. |
| **Halide tolerance factor $\tau$** | Bartel halide-perovskite stability descriptor | $\tau$: $\tau < 4.18$ | Used to reduce false positives where $t, \mu$ alone are insufficient. |

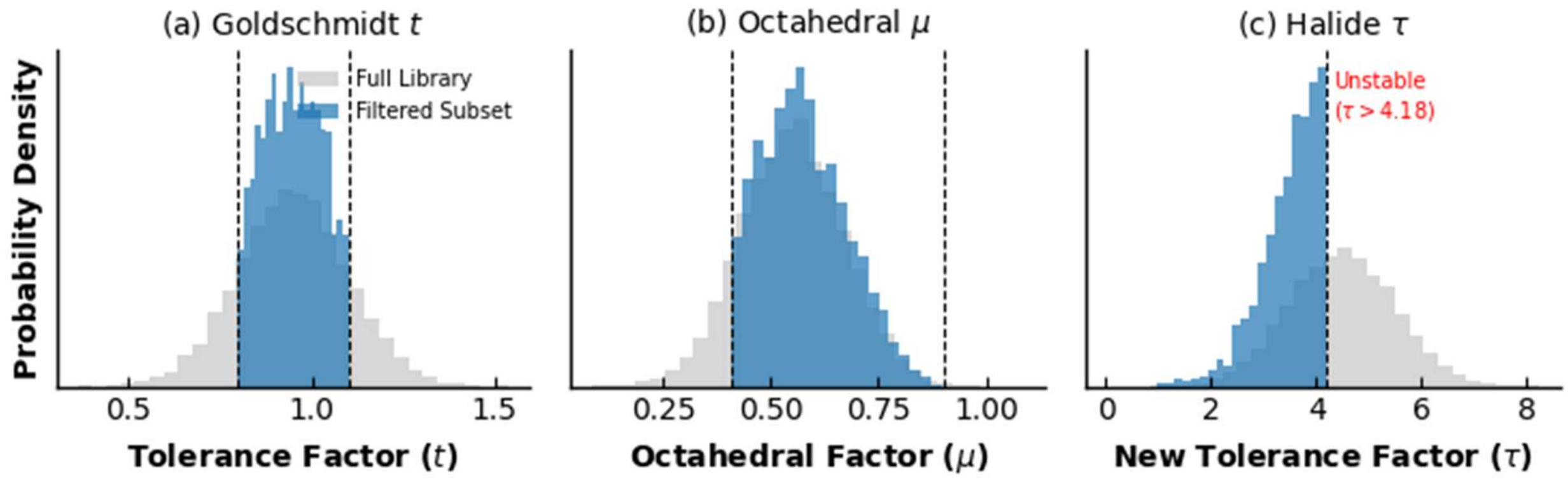

**Figure S4.** Distributions of geometric descriptors before and after filtering. Histograms of **(a)** Goldschmidt tolerance factor $\boldsymbol{t}$, **(b)** octahedral factor $\mu$, and **(c)** halide tolerance factor $\tau$ for the full enumerated $A_2BB'X_6$ library (gray, pre-filter) and the retained subset (blue, post-filter). The filters remove only the tails corresponding to severe size mismatch, while preserving a broad, physically plausible perovskite-formable region—supporting that the screening is systematic rather than cherry-picked. Vertical dashed lines indicate the acceptance thresholds listed in Table S8.